\documentclass[prx,aps,twocolumn,eqsecnum,superscriptaddress]{revtex4-1}
\usepackage{amssymb}
\usepackage{amsmath}
\usepackage{graphicx}
\usepackage[colorlinks=true,linkcolor=blue,anchorcolor=red,citecolor=blue,urlcolor=blue]{hyperref}
\usepackage{amsmath}

\begin{document}

\title{Quantum Transport in Topological Semimetals under Magnetic Fields (II)}

\author{Hai-Peng Sun}
\affiliation{Department of Physics, Harbin Institute of Technology, Harbin 150001, China}
\affiliation{Shenzhen Institute for Quantum Science and Engineering and Department of Physics, Southern University of Science and Technology, Shenzhen 518055, China}
\affiliation{Shenzhen Key Laboratory of Quantum Science and Engineering, Shenzhen 518055, China}

\author{Hai-Zhou Lu}
\email{Corresponding author: luhaizhou@gmail.com}
\affiliation{Shenzhen Institute for Quantum Science and Engineering and Department of Physics, Southern University of Science and Technology, Shenzhen 518055, China}
\affiliation{Shenzhen Key Laboratory of Quantum Science and Engineering, Shenzhen 518055, China}

\date{\today }

\begin{abstract}
We review our recent works on the quantum transport, mainly in topological semimetals and also in topological insulators, organized according to the strength of the magnetic field. At weak magnetic fields, we explain the negative magnetoresistance in topological semimetals and topological insulators by using the semiclassical equations of motion with the nontrivial Berry curvature. We show that the negative magnetoresistance can exist without the chiral anomaly. At strong magnetic fields, we establish theories for the quantum oscillations in topological Weyl, Dirac, and nodal-line semimetals. We propose a new mechanism of 3D quantum Hall effect, via the ``wormhole" tunneling through the Weyl orbit formed by the Fermi arcs and Weyl nodes in topological semimetals. In the quantum limit at extremely strong magnetic fields, we find that an unexpected Hall resistance reversal can be understood in terms of the Weyl fermion annihilation. Additionally, in parallel magnetic fields, longitudinal resistance dips in the quantum limit can serve as signatures for topological insulators.
\end{abstract}

\maketitle



\tableofcontents

\section{Introduction}

In the previous article, we have reviewed our recent works on quantum transport phenomena in topological Weyl and Dirac semimetals under magnetic fields \cite{Lu17fop}, which covers weak (anti-)localization \cite{Lu15prb-WAL,Dai16prbrc}, negative magnetoresistance in Weyl/Dirac semimetals \cite{LiH16nc}, and magnetotransport in the quantum limit \cite{Lu15prb-QL,ZhangSB16njp}. In this review, we summarize our recent works that were not addressed In Ref. \cite{Lu17fop}, specifically, on the 3D quantum Hall effect \cite{WangCM17prl}, quantum oscillations in Weyl/Dirac \cite{WangCM16prl} and nodal-line semimetals \cite{LiC18prl}, and negative magnetoresistance in topological insulators \cite{Dai17prl}, Weyl-node annihilation, \cite{ZhangCL17np} and vanishing backscattering \cite{ChenYY18prl} in the quantum limit. Following the structure in the previous review, these phenomena are organized depending on the strength of the magnetic field.
There are several recent review papers on topological semimetals \cite{WengHM16jpcm,YanBH17arcmp,Armitage18rmp,WangHC18cpb}.

The structure of this review is as the following.
In Sec. \ref{Sec:Models}, we introduce the models we used for describing Weyl, Dirac, nodel-line semimetals, and topological insulators.
In Sec. \ref{Sec:NMR}, we survey the negative magnetoresistance without chiral anomaly in topological insulators \cite{Dai17prl}.
In Sec. \ref{Sec:QO-Weyl}, we discuss quantum oscillations with the anomalous phase shift in topological semimetals \cite{WangCM16prl}.
In Sec. \ref{Sec:QO-NL}, we present the rules for phase shifts of quantum oscillations in topological nodal-line semimetals \cite{LiC18prl}.
In Sec. \ref{Sec:3DQHE}, we predict a 3D quantum Hall effect of Fermi arcs in topological semimetals \cite{WangCM17prl}.
In Sec. \ref{Sec:Annihilation}, we present the theory for the Weyl fermion annihilation in the quantum limit. The signature is recently observed as a sharp reversal in the Hall resistance of the topological Weyl semimetal TaP at extremely strong magnetic fields \cite{ZhangCL17np}.
In Sec. \ref{Sec:TI-QL}, we propose that resistance dips in the quantum limit can serve as a signature for topological insulators because of forbidden backscattering \cite{ChenYY18prl}.
Finally, we remark on future works in \ref{Sec:remark}.

\section{Models for topological semimetals and insulators}\label{Sec:Models}

In this section, we introduce the models we used for describing Weyl, Dirac, nodel-line semimetals, and topological insulators.

\subsection{Topological Weyl semimetal}

The topological Weyl and Dirac semimetal are  3D topological states of matter \cite{Wan11prb,Yang11prb,Burkov11prl,Xu11prl,Delplace12epl,Jiang12pra,Young12prl,Wang12prb,Singh12prb,Wang13prb,LiuJP14prb,Bulmash14prb}. Their energy
bands touch at the Weyl nodes which host monopoles. The Weyl nodes have
been verified in the Dirac semimetals Na$_3$Bi \cite{Wang12prb,Liu14sci,Xu15sci} and Cd$_3$As$_2$ \cite{Wang13prb,Liu14natmat,Neupane14nc,Borisenko14prl,Yi14srep,ZhangC17nc,LiCZ15nc,LiH16nc}, and the Weyl semimetal TaAs family  \cite{Huang15nc,Weng15prx,Lv15prx,Lv15np,Lv15prl,Xu15sci-TaAs,Yang15np,Liu15nmat,ZhangCL16nc,HuangXC15prx,Yang15np, Xu15np-NbAs,Xu16nc-TaP,Xu2016prl-TaAs,Belopolski2016prl,Belopolski2016prb-TaAs,Xu2015sciadv}, {\color{blue} TaIrTe$_4$ \cite{Belopolski2017nc} } and YbMnBi$_2$ \cite{Borisenko15arXiv}. Also, Weyl semimetals can be induced from half-Heusler compounds by applying magnetic fields \cite{Hirschberger16nmat,Felser16nmat,Shekhar18pnas}.

According to Sec. 2.1 of \cite{Lu17fop}, a Weyl semimetal can be described by a two-node Hamiltonian \cite{Shen17book,Okugawa14prb,Lu15Weyl-shortrange}
\begin{equation}\label{OC-Ham}
 H =A(k_x\sigma_x+k_y\sigma_y)+M (  k_w^2-\mathbf{k}^2)\sigma_z,
\end{equation}
where $ (\sigma_x,\sigma_y,\sigma_z)$ are the Pauli matrices, $\mathbf{k}=(k_x,k_y,k_z)$ is the wave vector , $A$, $M$, and $k_w$ are model parameters.
The eigen energies  of the Hamiltonian are $E_{\pm }^{\mathbf{k}}=\pm [M^2(k_w^2-\mathbf{k}^2)^2 +A^2(k_x^2+k_y^2)]^{1/2}$, with $+$ for the conduction band and $-$ for the valence band. The two Weyl nodes are at $(0,0,\pm k_w)$, and it has been demonstrated that the model is of  all the topological semimetal properties \cite{Lu15Weyl-shortrange}, in particular, the Fermi arcs \cite{ZhangSB16njp}, different from the $k\cdot \sigma$ model \cite{Ashby14epjb,Gorbar14prb,Lu15prb-WAL}. This model has the topological properties because of the $\sigma_z$ term \cite{Lu10prb}. For the 3D quantum Hall effect in Sec. \ref{Sec:3DQHE}, the above model is added with two trivial $D$ terms
\cite{Shen17book,Okugawa14prb,Lu15Weyl-shortrange}
\begin{equation}\label{QH-Ham}
 H=D_1k_y^2+D_2(k_x^2+k_z^2)+A(k_x\sigma_x+k_y\sigma_y)+M(k_w^2-\mathbf{k}^2)\sigma_z.
\end{equation}

\subsection{Topological Dirac semimetal}

A Dirac semimetal can be regarded as a Weyl semimetal and its time-reversal partner.
Dirac semimetals can be studied by using the Hamiltonian \cite{Wang12prb,Wang13prb,Jeon14nmat}
\begin{eqnarray}\label{Eq:H-Wang}
  H&=&\varepsilon_0(\mathbf{k})+
                      \begin{bmatrix}
                        M( \mathbf{k}) & Ak_+ & 0 & 0 \\
                        Ak_- & -M(\mathbf{k}) & 0 & 0 \\
                        0 & 0 & M(\mathbf{k}) & -Ak_- \\
                        0 & 0 & -Ak_+ & -M(\mathbf{k}) \\
                      \end{bmatrix}
                      \nonumber\\
&&  +\frac{\mu_B}{2}({\bm \sigma}\cdot\bm B)\otimes
                                                   \begin{bmatrix}
                                                     g_s & 0 \\
                                                     0 & g_p \\
                                                   \end{bmatrix},
\end{eqnarray}
where  $g_s$  is the $g$ factor for the  $s$ band,   $g_p$ is  the $g$ factor for the $p$ band \cite{Jeon14nmat}, $k_\pm=k_x\pm ik_y$,
$  \varepsilon_0(\mathbf{k})=C_0+C_1k_z^2+C_2(k_x^2+k_y^2)$, and $  M(\mathbf{k})=M_0+M_1k_z^2+M_2(k_x^2+k_y^2)$. The $x$, $y$, and $z$ axes in the Hamiltonian are defined along the [100], [010], and [001] crystallographic directions, respectively.

\subsection{Topological nodal-line semimetal}

Nodal-line semimetals \cite{Burkov11prb,Chiu14prb,Fang16cpb,Yang17prb}, in which the cross sections of  conduction and valence bands are closed rings [Fig. \ref{fig:torus}(a)] in momentum space \cite{Chen15nl,Bzduvsek16nature,FengX18prm,YiC18prb}, have the drumhead surface states, of which the direct
evidence is still missing \cite{ChenW18prl}. Recently, a new type of surface state called the floating band was discovered in the nodal-line semimetal ZrSiSe \cite{Zhu2018nc}. Besides, when the symmetries are  broken, the nodal-line semimetals may develop into Dirac semimetals, topological insulators, and surface Chern insulators \cite{ChenW18arXiv}. The nodal lines are predicted in many materials, such as HgCr$_2$Se$_4$ \cite{Xu11prl}, graphene networks \cite{Weng15prb}, Cu$_3$(Pd/Zn)N \cite{Yu15prl,Kim15prl}, SrIrO$_3$ \cite{Fang15prb,Chen15nc}, TlTaSe$_2$ \cite{Bian16prb}, Ca$_3$P$_2$ \cite{Xie15aplm,Chan16prb}, CaTe \cite{Du17npjqm}, compressed black phosphorus \cite{Zhao16prb}, CaAg(P/As) \cite{Yamakage15jpsj}, CaP$_3$ family \cite{Xu17prb}, PdS monolayer \cite{Jin17nanoscale}, Zintl compounds \cite{Zhu16prb}, Ba{\it MX}$_3$ ({\it M}=V, Nb and Ta, {\it X}=S, Se) \cite{Liang16prb}, rare earth monopnictides \cite{Zeng15arXiv}, alkaline-earth compounds \cite{Hirayama17nc,Huang16prb,Li16prl}, other carbon-based materials \cite{Chen15nl,WangJT16prl}, and metallic rutile oxides {\it X}O$_2$ ({\it X}=Ir, Os, Rd) \cite{SunY17prb}. So far, the nodal lines have been verified in ZrSiS \cite{Schoop16nc,Neupane16prb,ChenC17prb}, PbTaSe$_2$ \cite{Bian16nc,Chang2016prb},  InBi \cite{Ekahana17njp}  and PtSn$_4$ \cite{Wu16np}  by ARPES.


The Hamiltonian of nodal-line semimetals can be described as \cite{Bian16prb,Bian16nc}
\begin{equation}\label{eq:NL-model}
  H=\left\{\left[\hbar^2( k_x^2+ k_y^2)/2 m-u\right]\tau _3+\lambda  k_z \tau _1\right\}\otimes \sigma _0,
\end{equation}
where  ${\bf k}=(k_x,k_y,k_z)$ is the wave vector, $\tau$, $\sigma$ are the Pauli matrices,  $\lambda$, $m$, and $u$ are model parameters \cite{Bian16nc,Bian16prb}. The eigen energies  of the Hamiltonian are
$E_{\pm}=\pm \sqrt{[{\hbar^2(k_x^2+k_y^2)}/{2 m}-u]^2+\lambda^2k_z^2}$. When  $u$ is positive,  two bands intersect at zero energy as $k_x^2+k_y^2=2mu/\hbar^2$, which describes the nodal-line ring. The radius of the nodal ring is $\sqrt{2mu}/\hbar$ [Fig. \ref{fig:torus}(a)]. When  $E_F<u$, the dispersions result in a torus Fermi surface [Fig. \ref{fig:torus}(a)]. When $E_F>u$, the Fermi surface evolves into a drum-like structure [Fig. \ref{fig:torus}(a)]. Because of the low carrier density of the samples in experiments \cite{Schoop16nc,Neupane16prb}, we focus on  the case that  $E_F<u$. Moreover, the model in Eq. (\ref{eq:NL-model}) is of the mirror reflection symmetry  \cite{Bian16prb,Bian16nc}. Nodal lines can also be protected by other symmetries \cite{Fang16cpb,Kim15prl,Bzduvsek16nature,Alexandradinata18prbrc}, for example, two-fold screw rotation \cite{Fang15prb,Chen15nc}, four-fold rotation, inversion \cite{Fang16cpb} and non-symmorphic symmetry through a glide plane \cite{Schoop16nc,Neupane16prb}, etc.

\subsection{Topological insulator}

3D topological insulators can be described by the $k\cdot p$ Hamiltonian \cite{Zhang09np,Shen17book,Nechaev16prb}
\begin{equation}\label{full_hamiltonian}
H_0=C_{\mathbf k}+
\begin{pmatrix}
M_{\mathbf k} & 0 & i V_n k_z & -i V_{\perp} k_-\\
0 & M_{\mathbf k} & i V_{\perp} k_+ & i V_n k_z\\
-i V_n k_z & -i V_{\perp} k_- & -M_{\mathbf{k}} & 0\\
i V_{\perp} k_+ & -i V_n k_z & 0 & -M_{\mathbf{k}}
\end{pmatrix},
\end{equation}
where $M_{\mathbf{k}}=M_0+M_{\perp}(k_x^2+k_y^2)+M_z k_z^2$, $C_{\mathbf{k}}=C_0+C_{\perp}(k_x^2+k_y^2)+C_z k_z^2$,
 $C_i$, $M_i$ and $V_i$ are model parameters.
The model depicts a 3D strong topological insulator as $M_0M_\perp<0$ and $M_0M_z<0$ \cite{Shen17book}.
There are four energy bands $\varepsilon_n(\mathbf{k})$ near the $\Gamma$ point, two conduction bands and two valence bands (see Fig. 2 In Ref. \cite{Dai17prl}). The model has been shown effective to give proper descriptions for the topological surface states \cite{Lu10prb,Shan10njp,ZhangY10np} and explain the negative magnetoresistance in topological insulators \cite{Dai17prl,Wang12nr,He13apl,Wiedmann16prb,Wang15ns}.

In the presence of the magnetic field, the Zeeman Hamiltonian reads
\begin{eqnarray}\label{zeeman}
    H_Z=\frac{\mu_B}{2}\left(
    \begin{array}{cccc}
        g_{z}^{v}B_z  & g_{p}^vB_- & 0 & 0\\
        g_{p}^v B_{+} & -g_{z}^v B_z & 0 & 0\\
        0 & 0 &  g_{z}^c B_z & g_{p}^c B_- \\
        0 & 0 & g_{p}^c B_+ & -g_{z}^c B_z
    \end{array}
    \right),
\end{eqnarray}
where $g_{v/c,z/p}$ are Land\'{e} g-factors for valence/conduction bands along the $z$ direction and in the $x-y$ plane and $\mu_B$ is the Bohr magneton.

\section{Weak field: Negative magnetoresistance in topological insulators}\label{Sec:NMR}

\begin{figure}[htpb]
\centering
\includegraphics[width=0.9\columnwidth]{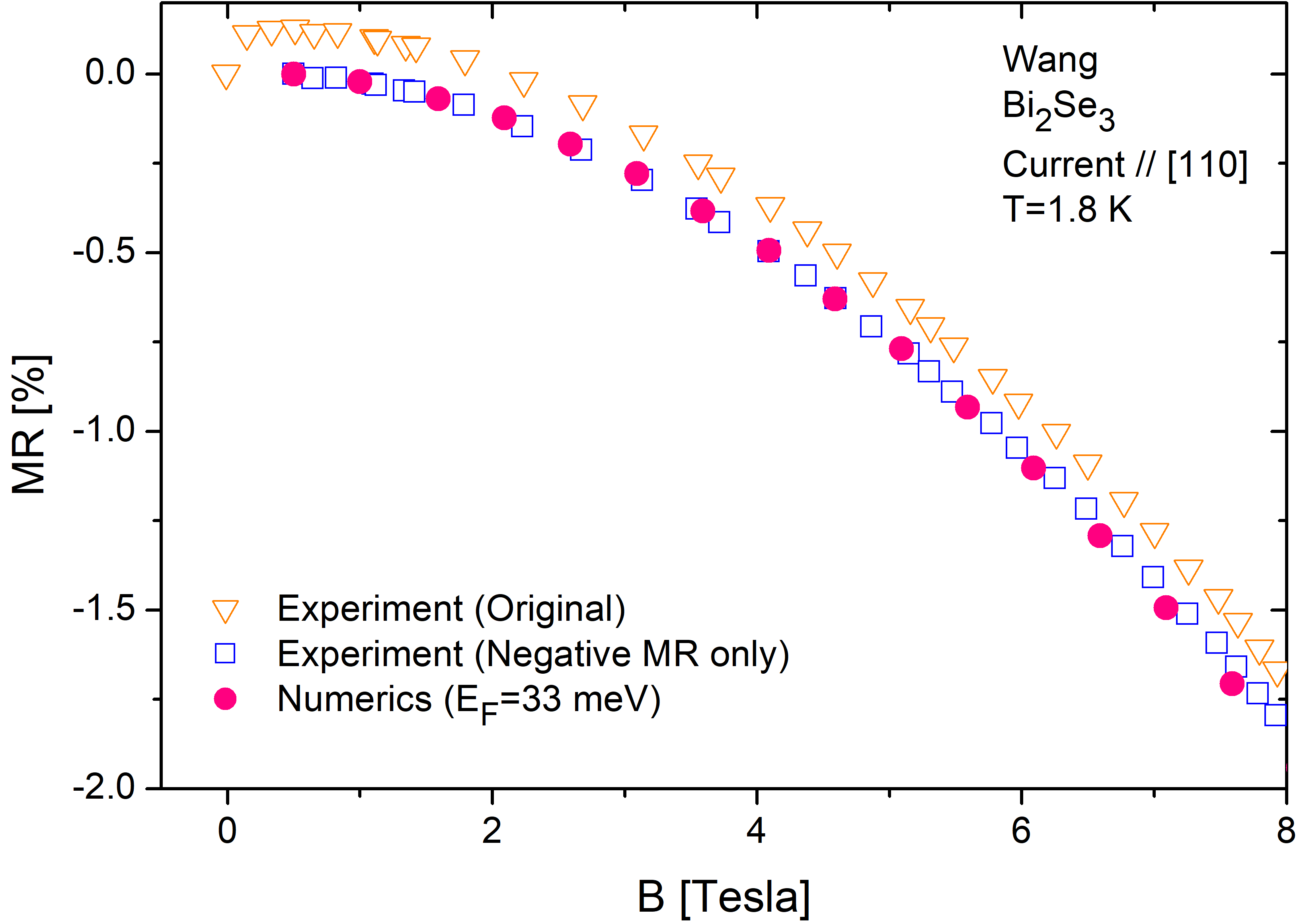}
\includegraphics[width=0.9\columnwidth]{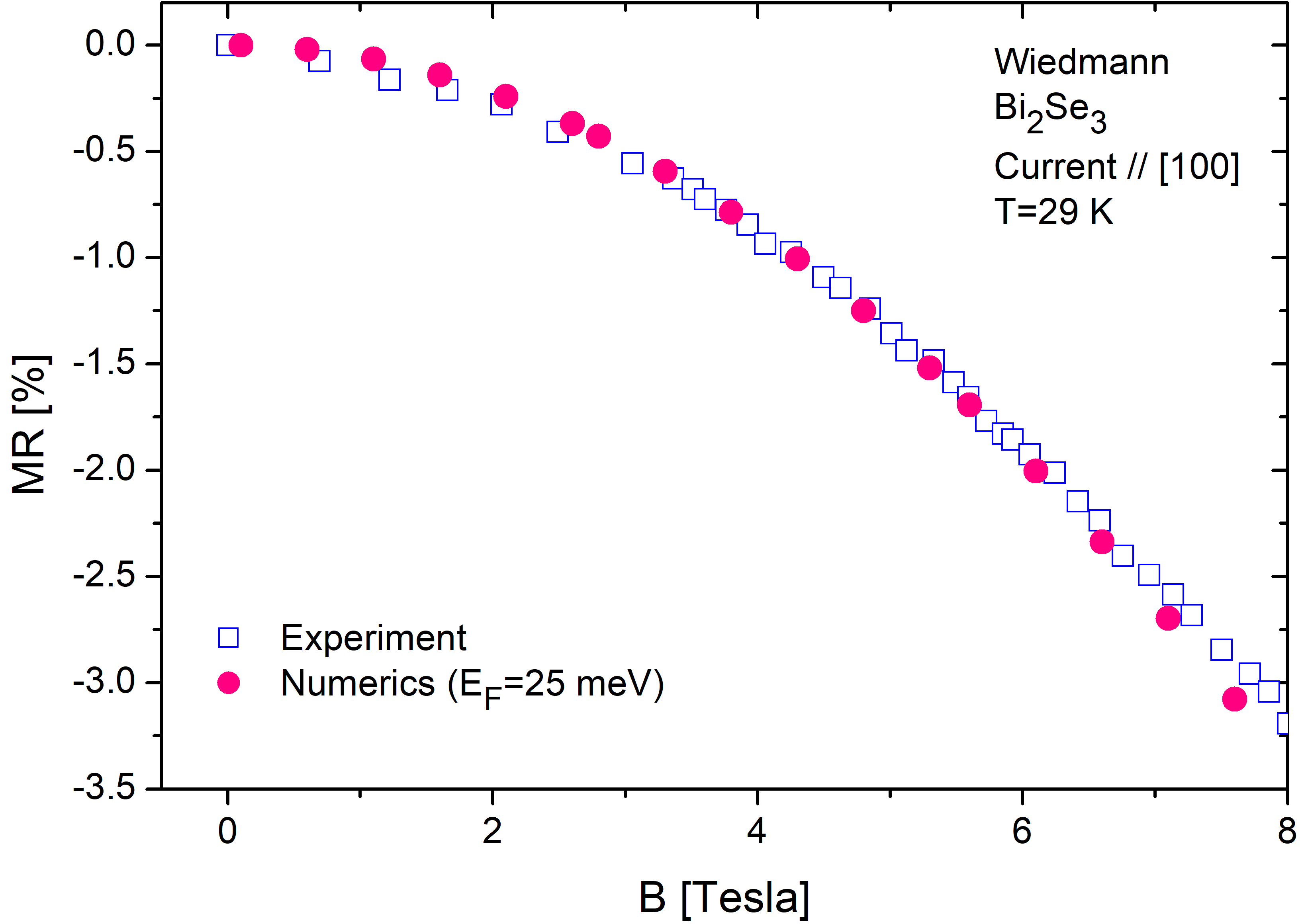}
\includegraphics[width=0.9\columnwidth]{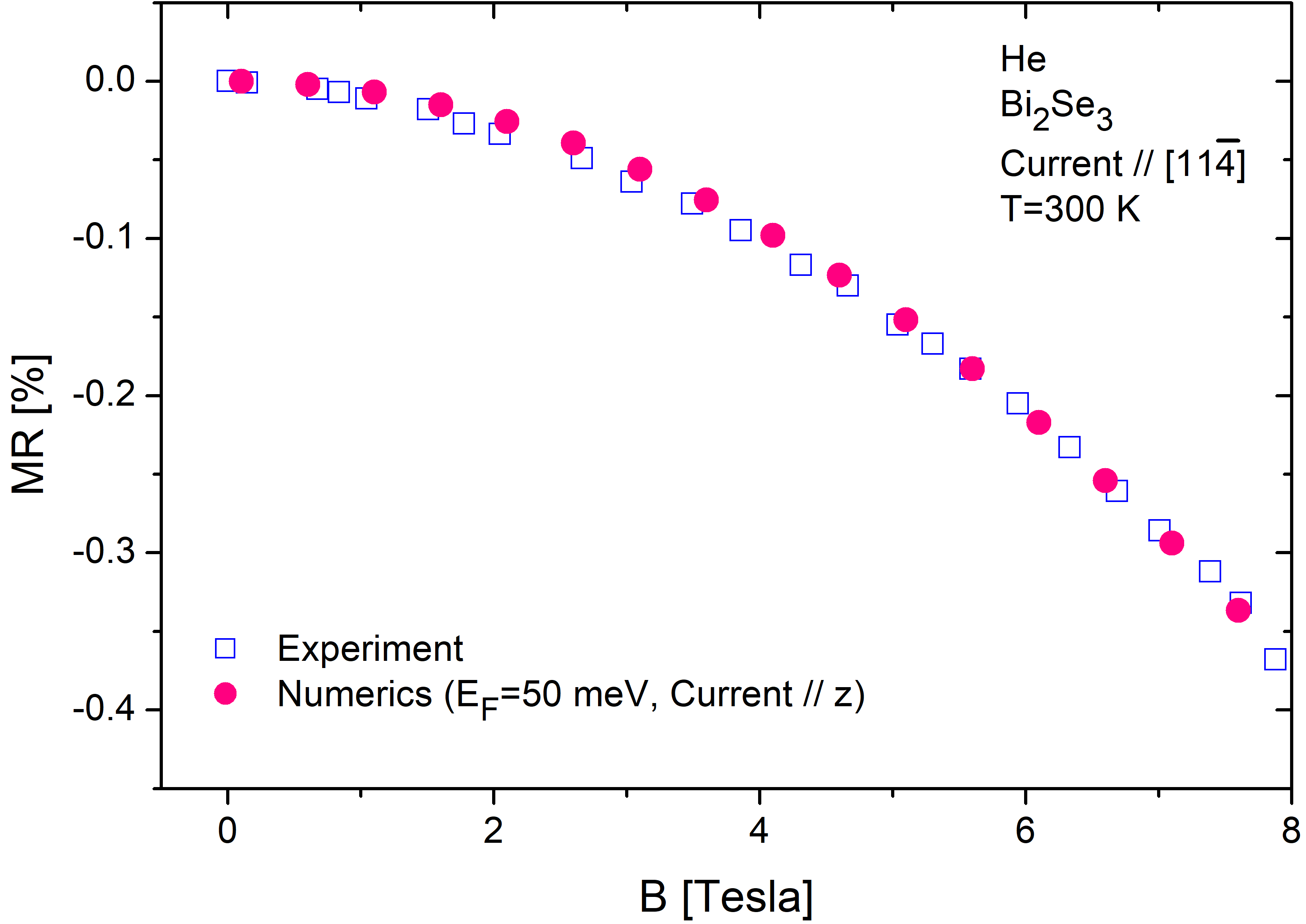}
\caption{Adapted from Ref. \cite{Dai17prl}. The comparison between the calculated negative magnetoresistance \cite{Dai17prl} and the experiments \cite{Wang12nr,Wiedmann16prb,He13apl}. The Fermi energy $E_F$ is measured from the conduction band bottom. $E_F$ falls in a reasonable range.  $x,y,z$ in the model \eqref{full_hamiltonian} correspond to [100], [010], and [001] crystallographic directions, respectively. The numerically calculated magnetoresistance along the $z$ axis is used to approach the experimental magnetoresistance along the $[11\bar{4}]$ direction by He \emph{et al.}.
Other model parameters are from the $k\cdot p$ calculations \cite{Nechaev16prb} and experiments \cite{Wolos16prb}, $M_0$=-0.169 eV, $M_z$=3.351 eV$\mbox{\AA}^2$, $M_{\perp}$=29.36 eV $\mbox{\AA}^2$, $V_{\perp}$=2.512 eV$\mbox{\AA}$, $V_n$=1.853 eV$\mbox{\AA}$, $C_0$=0.048 eV, $C_z$=1.409 eV $\mbox{\AA}^2$, $C_{\perp}$=13.9 eV$\mbox{\AA}^2$, the g-factors $g_{z}^v=g_{z}^c=30$ and $g_{p}^v=g_{p}^c=-20$. }
\label{Fig:comparison}
\end{figure}

Recently discovered topological semimetals can host the chiral
anomaly, namely, the violation of the conservation of chiral current \cite{Adler69pr,Bell69Jackiw,Nielsen81npb}, which is widely believed to be the cause of the negative magnetoresistance \cite{Kim13prl,Kim14prb,Li16np,ZhangCL16nc,HuangXC15prx,Xiong15sci,LiCZ15nc,ZhangC17nc,LiH16nc,Arnold16nc,YangXJ15arXiv,YangXJ15arXiv-NbAs,WangHC16prb}. Nevertheless, in topological insulators, where the chiral anomaly is not well defined in the momentum space, a negative magnetoresistance can also be observed. This results in great confusion \cite{Wang12nr,He13apl,Wiedmann16prb,Wang15ns,Breunig17nc,Assaf17prl,ZhangMH18an} on the explanation of the negative magnetoresistance. Lately, it is found that the chiral anomaly in real space can be defined in a quantum spin Hall insulator \cite{Fleckenstein2016prb}.
In Ref. \cite{Dai17prl}, we use the semiclassical Boltzmann formalism with the Berry curvature and orbital moment, to explain the negative magnetoresistance in topological insulators, and show a quantitative agreement with the experiments (see Fig. \ref{Fig:comparison}).

\subsection{Anomalous velocity}

The Berry curvature and orbital moment will induce  the anomalous velocity, which may lead to the negative magnetoresistance. In experiments, the negative magnetoresistance exists above $T=100$ K~\cite{Wiedmann16prb},
thus quantum interference mechanisms can be excluded.
In addition, due to the poor mobility in the topological insulators Bi$_2$Te$_3$ and Bi$_2$Se$_3$ \cite{Culcer12pe}, when the magnetic field is  up to 6 Tesla, the Landau levels cannot
be well-formed.
In the semiclassical regime, the electronic transport can be described by the equations of motion \cite{Sundaram99prb}
\begin{eqnarray}\label{EOM-maintext}
\dot{\mathbf{r}}=\frac{1}{\hbar}\nabla_{\mathbf{k}}\widetilde{\varepsilon}_{\mathbf{k}}
-\dot{\mathbf{k}}\times \mathbf{\Omega}_{\mathbf{k}}, \ \ \ \
\dot{\mathbf{k}}=-\frac{e}{\hbar}(\mathbf{E}+\dot{\mathbf{r}}\times\mathbf{B}),
\end{eqnarray}
where both the position $\mathbf{r}$
and wave vector $\mathbf{k}$ appear simultaneously,
$\dot{\mathbf{r}}$ and $\dot{\mathbf{k}}$ are their time derivatives, $-e$ is the electron charge,
$\mathbf{E}$ and $\mathbf{B}$ are external electric and magnetic fields,
respectively.  $\widetilde{\varepsilon}_{\mathbf{k}}=\varepsilon_{\mathbf{k}}-\mathbf{m} \cdot \mathbf{B}$,
$\varepsilon_{\mathbf{k}}$ is the band dispersion, $\mathbf{m}$ is the orbital moment
induced by the semiclassical self-rotation of the Bloch wave packet, and $\mathbf{\Omega}_{\mathbf{k}}$ is the Berry curvature \cite{Xiao10rmp}.
In the linear-response limit ($\mathbf{E}=0$), Eq.~\eqref{EOM-maintext} yields an effective velocity
\begin{eqnarray}\label{Eq:velocity}
&&\dot{\mathbf{r}}= [\widetilde{\mathbf{v}}_{\mathbf{k}}
+(e/\hbar)\mathbf{B}(\widetilde{\mathbf{v}}_{\mathbf{k}}\cdot\mathbf{\Omega}_{\mathbf{k}})]/D_{\mathbf{k}},
\end{eqnarray}
where $D_{\mathbf{k}}^{-1}$ is the correction to the density of states, and
\begin{equation}\label{v_and_d}
\begin{split}
\widetilde{\mathbf{v}}_{\mathbf{k}}=\mathbf{v}_{\mathbf{k}}-\frac{1}{\hbar}
\nabla_\mathbf{k} (\mathbf{m}_{\mathbf{k}}\cdot\mathbf{B}),\ \ \
D_{\mathbf{k}}=1+\frac{e}{\hbar}\mathbf{B}\cdot\mathbf{\Omega}_{\mathbf{k}}.
\end{split}
\end{equation}
Due to the Berry curvature, the velocity develops an anomalous term, which is proportional to $\mathbf{B}$. Note that the conductivity is the current-current (velocity-velocity) correlation \cite{Mahan1990},
thus the presence of the anomalous velocity is expected to generate an extra conductivity, which grows with the magnetic field, namely, a negative magnetoresistance.
It has been implied
that the negative magnetoresistance in topological semimetals is related to the Berry curvature \cite{Son13prb,Yip15arXiv,Lu17fop}. In our previous review \cite{Lu17fop}, we have shown that the Berry curvature \cite{Sundaram99prb} can lead to a conductivity correction that grows with magnetic field $B$,
\begin{eqnarray}
\delta \sigma (B) \propto \frac{B^2}{k_F^2}\propto \frac{B^2}{n^{2/3}},
\end{eqnarray}
where $k_F$ is the Fermi wave vector and the carrier density $n$ is directly proportional to $k_F^3$. This is an alternative understanding to the negative magnetoresistance induced by the chiral anomaly.
Ref. \cite{Dai17prl} shows that this mechanism is large enough in topological insulators as those observed in the experiments, where the relative magnetoresistance
can exceed -1\% in a parallel magnetic field of several Tesla \cite{Wang12nr,He13apl,Wang15ns,Wiedmann16prb,Breunig17nc,Assaf17prl,ZhangMH18an}.

\subsection{Magnetoresistance formula}

In our calculation, the relative magnetoresistance is defined as
$\mathrm{MR}_{\mu}(B_\mu) =[1/\sigma^{\mu\mu}(B_\mu)-1/\sigma^{\mu\mu}(0)]/[1/\sigma^{\mu\mu}(0)]$. In the semiclassical Boltzmann formalism, the longitudinal conductivity $\sigma^{\mu\mu}$ is contributed by all the bands crossing the Fermi energy, and for band $n$ \cite{Yip15arXiv}
\begin{equation}
\sigma^{\mu\mu}\label{conduction_tensor}
=\int\frac{d^{3}\mathbf{k}}{(2\pi)^{3}}
\frac{e^{2}\tau}{D_{\mathbf k}}
\left(\widetilde{v}^{\mu}_{\mathbf{k}}
+\frac{e}{\hbar}B^{\mu}\widetilde{v}^{\nu}_{\mathbf{k}}\Omega^{\nu}_{\mathbf{k}}\right)^2
\left(-\frac{\partial\widetilde{f}_0}{\partial\widetilde{\varepsilon}}\right),
\end{equation}
where $n$ is suppressed for simplicity, $D_{\mathbf{k}}$ and $\widetilde{v}^{\mu}_{\mathbf{k}}$ are
given by Eq.~\eqref{v_and_d}, $\widetilde{f}_0$ is the Fermi distribution in equilibrium,
the transport time $\tau$ is assumed to be a constant in
the semiclassical limit \cite{Burkov14prl-chiral}.
For the $n$-th band of the Hamiltonian $H$,
the $\xi$ component of the Berry curvature
vector can be found as
$\Omega_{n\mathbf{k}}^{\xi}=\Omega_{n\mathbf{k}}^{\mu\nu}\varepsilon_{\mu\nu\xi}$,
where $\xi,\mu,\nu$ stand for
$x,y,z$, and $\varepsilon_{\mu\nu\xi}$ is the Levi-Civita anti-symmetric tensor, and
\begin{eqnarray} \label{berry}
\Omega_{n\mathbf{k}}^{\mu\nu}  = -2\sum_{n'\neq n}
\frac{\mathrm{Im}\langle n|
\partial H/\partial k_{\mu} |n'\rangle \langle n'|
  \partial H/\partial k_{\nu} |n\rangle}{(\varepsilon_n-\varepsilon_n')^2},
\end{eqnarray}
where $H=H_0+H_Z$. The orbital moment $\mathbf{m}$ can be found as
\begin{equation}\label{orbital}
m_{n\mathbf{k}}^{\mu\nu}  = -\frac{e}{\hbar}\sum_{n'\neq n}
\frac{\mathrm{Im}\langle n|
\partial H/\partial k_{\mu}  |n'\rangle \langle n'|
  \partial H /\partial k_{\nu} |n\rangle}
{\varepsilon_n-\varepsilon_n'}.
\end{equation}
The Zeeman energy can induce a finite distribution of $\mathbf{\Omega}$ and $\mathbf{m}$ \cite{Dai17prl}.

\subsection{Comparison with negative magnetoresistance in experiments}

Figures~\ref{Fig:comparison} shows that the numerically calculated relative magnetoresistance in parallel magnetic fields are negative and decrease
monotonically with the magnetic field.
They can be fitted by $-B^2$ at small magnetic fields and conform to the Onsager reciprocity MR$(B)$=MR$(-B)$. We also use a tight-binding model \cite{Mao11prb} to justify the calculation. Figure \ref{Fig:comparison} shows a good agreement on the negative magnetoresistance between the experiments and our numerical calculations. The current direction and temperature are from the experiments and the model parameters are from the $k\cdot p$ calculations \cite{Nechaev16prb} and experiments \cite{Wolos16prb}. In the experiment by Wang \emph{et al}. \cite{Wang12nr}, the temperature is 1.8 K, so the original data (orange triangles) has a positive magnetoresistance near zero field due to the weak anti-localization \cite{Checkelsky09prl,Chen10prl,Wang11prb,He11prl}. The weak anti-localization induces a positive magnetoresistance \cite{Lu15prb-WAL}, which is subtracted before the comparison. In the experiments by Wiedmann \emph{et al}. \cite{Wiedmann16prb} and He \emph{et al}. \cite{He13apl}, the temperatures are 29 K and 300 K, there is no weak anti-localization effect.
The negative magnetoresistance does not change much with temperature \cite{Dai17prl}, consistent with the experiments and showing the semi-classical nature of the negative magnetoresistance. The negative magnetoresistance becomes enhanced as
the Fermi level approaches the band bottom \cite{Dai17prl}, indicating the role of the Berry curvature.
Ref. \cite{Dai17prl} also shows that the signs of $g$-factors in the Zeeman coupling determine
the signs of magnetoresistance qualitatively.
In the experiment, the techniques used for topological insulators,
for example, electron spin resonance and quantum oscillations,
can not determine the signs of $g$-factors but only their absolute values~\cite{Wolos16prb,Kohler75pssb}.
Transport measurements can determine
the sign of the $g$-factor only in specific setups \cite{Srinivasan16prbrc}.
The orbital moment has been neglected
in most of the literature
studying the magneto-transport using the semiclassical formalism \cite{Son13prb,Yip15arXiv}.
As shown in Refs.~\cite{Morimoto16prb,GaoY17prb},
the orbital moment is essential for the magnetoresistance anisotropy in a Weyl semimetal.
Moreover, the correction $\bf m\cdot \bf B$ to $\varepsilon(\bf k)$
can enhance the band separation and the negative magnetoresistance.
The orbital moment effectively enhances the MR$_x$ a few
times larger. MR$_z$ can be even positive without $\mathbf{m}$.

\subsection{Discussions}

The conventional equations of motion in the low-field semiclassical regime ($\omega \tau \ll 1$) are only accurate to the linear order in the external fields ($\mathbf{E}$ and $\mathbf{B}$), thus for studying magnetoconductivity which is an intrinsically nonlinear coefficient, the obtained results would not be complete. Based on a recently developed semiclassical theory with second-order accuracy \cite{GaoY14prl,GaoY15prb}, a complete theory of  magnetoconductivity for general 3D nonmagnetic metals was formulated within the Boltzmann framework with the relaxation time approximation \cite{GaoY17prb}. The work shows several surprising results. First, there is an important previously unknown Fermi surface contribution $\delta \sigma^{int}$ to the magnetoconductivity, termed as the intrinsic magnetoconductivity, because the ratio $ \delta \sigma^{int}/\sigma_0 $ is independent of the relaxation time. Here $\sigma_0$ is the conductivity at $B=0$. Second, a pronounced $\delta \sigma^{\rm int}$ term can lead to the violation of Kohler's rule. Previously, any deviation from Kohler's rule is usually interpreted as from factors beyond the semiclassical description or from the presence of multiple types of carriers or multiple scattering times. The result here reveals a new mechanism for the breakdown of Kohler's rule. Third, $\delta \sigma^{int}$ can lead to a positive longitudinal magnetoconductivity (or negative longitudinal magneto-resistivity). The effect is independent of chiral anomaly effect for the Weyl/Dirac fermions, and it can occur for a generic doped semiconductor without any band crossings. This indicates that positive longitudinal magnetoconductivity measured in the semiclassical regime alone cannot be regarded as smoking-gun evidence for the existence of topological band crossings. The intrinsic contribution generally exists in 3D metals with nontrivial Berry curvatures, and should be taken into account when interpreting experimental results. It may already play an important role behind the puzzling magnetotransport signals observed in recent experiments on TaAs$_2$ and related materials.

In the quantum limit where only the lowest Landau band is occupied, magnetoresistance depends subtly on scattering mechanisms \cite{Lu15prb-QL,Goswami15prb,ZhangSB16njp}, rather than the Berry curvature and orbital moment. The current-jetting effect is usually induced by inhomogeneous currents when attaching point contact electrodes to a large bulk crystal and may also hamper the interpretation of the negative magnetoresistance data \cite{dosReis16njp}. A recent work also has pointed out that the negative magnetoresistance may exist without the chiral anomaly \cite{Andreev18prl}. In (Bi$_{1-x}$In$_{x}$)$_2$Se$_3$, it is proposed that the in-plane negative magnetoresistance is due to the topological phase transition enhanced intersurface coupling near the topological critical point \cite{ZhangMH18an}. In addition, it is also found that the magnetoresistance is robust against the deviation from the ideal Weyl Hamiltonian, such as  the shifted Fermi energy, nonlinear dispersions, and the Weyl node annihilation \cite{Ishizuka18arXiv}.

\section{Strong field: Quantum oscillation in Weyl and Dirac semimetals}\label{Sec:QO-Weyl}

When applying a magnetic field in the $z$-direction, the energy spectrum evolves into a series of 1D Landau bands \cite{Lu15Weyl-shortrange,ZhangSB16njp} (see Fig. 1(b)-(c) of \cite{WangCM16prl}), which result in the Shubnikov-de Haas (SdH) oscillation of resistance. The oscillation of the resistivity $\rho$ can be demonstrated by the Lifshitz-Kosevich formula \cite{Shoenberg84book}
\begin{equation}\label{Eq:rho-cos}
  \rho \sim \cos[2\pi(F/B+\phi)],
\end{equation}
where $\phi$ is the phase shift, $F$ is the oscillation frequency and $B$ is the magnitude of magnetic field.  $F$ and $\phi$ can reveal valuable details on the Fermi surface of the material.

The phase shift of each frequency component can be argued as the following
\begin{eqnarray}\label{eq:phase}
\phi=-{1}/{2}+{\phi_{\rm B}}/{2\pi}+\phi_{\rm 3D},
\end{eqnarray}
where  $\phi_{\rm 3D}=\mp1/8$ is a correction, which emerges only in 3D, and  $\phi_{\it B}$ is the Berry phase \cite{Mikitik99prl,Xiao10rmp}. The curvature of the Fermi surface along the direction of the magnetic field determines the sign of $\phi_{\rm 3D}$ \cite{Lifshitz56jetp,Shoenberg62,Coleridge72jpf,Lukyanchuk04prl}.
When the cross section is maximum, for electron carriers  $\phi_{\rm 3D}=-1/8$ and for hole carriers  $\phi_{\rm 3D}=1/8$; when the cross section is minimum, for electron carriers $\phi_{\rm 3D}=1/8$ and for hole carriers $\phi_{\rm 3D}=-1/8$. For the sphere Fermi surface, there is only a maximum, thus for electron carriers $\phi_{\rm 3D}=-1/8$  and for hole carriers $\phi_{\rm 3D}=1/8$ . For a  parabolic energy band,  it does not have Berry phase, thus the phase shift is $\pm 1/2$ and $\pm 5/8$ in 2D and 3D, respectively. However, a linear energy band (for example, Weyl and Dirac semimetals\cite{Volovik03book,Wan11prb,Xu11prl,Burkov11prl,Yang11prb}) has an extra $\pi$ Berry phase \cite{Mikitik99prl}, thus the phase shift is $0$ \cite{ZhangYB05nat} and $\pm 1/8$ \cite{Murakawa13sci} in 2D and 3D, respectively.
In a nodal-line semimetal \cite{Burkov11prb,Chiu14prb,Fang16cpb,Yang17prb}, an electron can collect a nontrivial $\pi$ Berry phase around the loop encircling the nodal line \cite{Fang15prb}. The phase shifts of 2D and 3D bands with linear and parabolic dispersions  are summarized in Table \ref{tab:phase}.

\begin{table}[htbp]
\caption{ Adapted from Ref. \cite{LiC18prl}. Phase shifts $\phi$ in Eq. (\ref{Eq:rho-cos}) for systems with different dispersions (linear or parabolic) and dimensionalities (2D or 3D). $B_z$ and $B_{||}$ are the magnetic fields out of and in the nodal-line plane. $\alpha$, $\beta$, $\gamma$, and $\delta$ correspond to the cross sections of the torus Fermi surface in Fig. \ref{fig:torus}. }
\begin{ruledtabular}
\begin{tabular}{cccc}
System & Electron carrier & Hole carrier  \\
\hline
2D parabolic & $-1/2$  & 1/2
\\
3D parabolic & $-5/8$   & 5/8
\\
2D linear & 0  & 0
\\
3D linear  & $-1/8$ & 1/8
\\
Nodal-line in $B_z$ & $-5/8$ $(\alpha)$, $5/8$ $(\beta)$   & $5/8$ $(\alpha)$, $-5/8$ $(\beta)$
\\
Nodal-line in $B_{||}$ & $-5/8$ $(\gamma)$, $1/8$ $(\delta)$   & $5/8$ $(\gamma)$, $-1/8$ $(\delta)$
\tabularnewline
\end{tabular}\label{tab:phase}
\end{ruledtabular}
\end{table}

Topological Weyl/Dirac semimetals and nodal-line semimetals provide a novel platform to explore the nontrivial Berry phase \cite{Murakawa13sci,He14prl,Novak15prb,Zhao15prx,Du16scpma,YangXJ15arXiv-NbAs,WangZ16prb,Xiong15sci,Cao15nc,ZhangCL15arXiv-TaAs,Narayanan15prl,Park11prl,Xiang15prb,Tafti16np,Luo15prb,Dai16prbrc,Arnold16prl,Klotz16prb,dosReis16prb,Sergelius16srep}.
In Ref. \cite{WangCM16prl}, we show that, near the Lifshitz point, the phase shift of the quantum oscillation can go beyond recognized values of $\pm 1/8$ or $\pm 5/8$ and nonmonotonically move toward a wide range between $\pm 7/8$ and $\pm 9/8$. However, these values in experiments may be misunderstood as $\pm 1/8$. For Dirac semimetals, the total phase shift adopts the discrete values of  $\pm 1/8$ or $\pm 5/8$. In recent experiments of electron carriers,  the positive phase shifts are observed and may be explained by our findings. In addition, a new beating pattern, resulting from the topological band inversion, is found.
Up to now, quantum oscillations have been inspected experimentally in HfSiS \cite{Kumar17prb}, ZrSiS \cite{Singha17pnas,Ali16sa,WangX16aem,Lv16apl,Hu16prb,PanH18srep}, ZrSi(Se/Te) \cite{Hu16prl} and ZrGe(S/Se/Te) \cite{Hu17prb}, but the phase shifts have been concluded different. In Ref. \cite{LiC18prl},  the phase shifts and frequencies  (Table \ref{tab:freq-pha}) of nodal-line semimetals are extracted from analytic results of the calculated resistivity.  We also summarize the generic rules for phase shifts in random cases (Table \ref{tab:rule}). The generic rules assist us to handle several materials, for example,  ZrSiS  and Cu$_3$PdN \cite{Singha17pnas,Ali16sa,WangX16aem,Lv16apl,Hu16prb,PanH18srep}.

\subsection{Quantum oscillation in linear and parabolic limits of a Weyl semimetal}

The resistivity is calculated in two direction configurations according to linear response theory \cite{Charbonneau82jmp,Vasilopoulos84jmp, WangCM12prb,WangCM15prb}. For the longitudinal configuration, the resistivity $\rho_{zz}$ is examined along $z$ direction. For the transverse configuration, the resistivity $\rho_{xx}$ is examined along $x$ direction. The magnetoresistivity in the linear dispersion limit and parabolic dispersion limit takes the form of Eq. (\ref{Eq:rho-cos}). In Table \ref{tab:two-limits}, we list the analytic expressions for the phase shift $\phi$ and frequency $F$ in these two limits.

\begin{table}[htb]
    \centering
\caption{Adapted from Ref. \cite{WangCM16prl}. The analytical expressions for the frequency $F$ and phase shift $\phi$ in the resistivity formula Eq. (\ref{Eq:rho-cos}) in the linear and parabolic dispersion limits for electron carriers in a Weyl semimetal. We define $E_F'\equiv E_F+ Mk_w^2 $. }\label{tab:two-limits}
\begin{ruledtabular}
\begin{tabular}{ccccc}
                           & \multicolumn{2}{c}{Longitudinal $\rho_{zz}$} & \multicolumn{2}{c}{Transverse $\rho_{xx}$} \\
                     &Parabolic                 &Linear       &Parabolic     &Linear\\ \hline
\rule{0pt}{0.65cm}$F$      & $\hbar E_F'/2eM$       & $\hbar E_F^2/2eA^2$           &  $\hbar E_F'/2eM$            & $\hbar E_F^2/2eA^2$      \\
\rule{0pt}{0.65cm}$\phi$             & -5/8            & -1/8 & -5/8 & -1/8\\
\end{tabular}
\end{ruledtabular}
\end{table}

\subsection{Resistivity peaks and integer Landau indices}

In the experiment, the peak positions or valley positions, that are on the $B$ axis, are given the integer Landau indices $n$, then the phase shift $\phi$ and frequency $F$ can be extracted from a plot of $n$ and $1/B$ [see inset of Fig. 1(d) of \cite{WangCM16prl}]. Nevertheless, it is still in debate that whether the peaks \cite{Murakawa13sci,Zhao15prx,He14prl,Zhao15prx,Du16scpma,YangXJ15arXiv-NbAs,WangZ16prb,Cao15nc} or valleys \cite{Narayanan15prl,Qu10sci,Park11prl,Luo15prb} should be given the Landau indices. Our results explicitly reveal that the resistivity peaks of  $\rho_{zz}$ and $\rho_{xx}$ emerge near the Landau band edges and are in correspondence  with the integer Landau indices.
We evaluate the resistivity components theoretically from the conductivity components \cite{Datta1997,Vasko06book}. For the longitudinal configuration, the resistivity $\rho_{zz}$=$1/\sigma_{zz}$, where the conductivity  $\sigma_{zz}$ is along $z$ direction. Near band edges, the conductivity $\sigma_{zz}$ shows valleys due to vanishing velocities, thus $\rho_{zz}$ shows peaks.
For the transverse configuration, $\rho_{xx}=\sigma_{yy}/(\sigma_{yy}^2+\sigma_{xy}^2)$, and the longitudinal Hall conductivity and field-induced Hall conductivity are as follows
\begin{eqnarray}\label{sigma-yx-delta}
\sigma_{yy}= \frac{\sigma_0(1+\delta)}{1+(\mu B)^2},\ \  \sigma_{yx} = \frac{ \mu B \sigma_0 }{1+(\mu B)^2}\left[1-\frac{\delta}{(\mu B)^2} \right], \nonumber \\
\end{eqnarray}
where $\delta\ll 1$ is the oscillation part and $\sigma_0$ represents the zero-field conductivity.  In $\sigma_{yx}$, the $\delta$ term, which comes from the disorder scattering, was rarely considered before. A consequence of the $\delta$ term is that $\rho_{xx} \approx  (1+\delta)/\sigma_0 $, up to the first order of $\delta$. As $\rho_{xx}$ and $\sigma_{yy}$ are both proportional to $1+\delta$, their peaks are lined up for the random ratio of $\sigma_{yx}$ to $\sigma_{yy}$ (but when $\sigma_{yx}\ll \sigma_{yy}$, the oscillation is so weak that it is hard to be observed ). This new finding comes from the disorder scattering term $\delta$ in the Hall conductance. At the same time, the $\sigma_{zz}$ valleys are lined up with the $\sigma_{yy}$ peaks, since $\sigma_{zz}$, which comes from diffusion, is in proportion to the scattering times. However, $\sigma_{yy}$, which arises from hopping is inversely in proportion to the scattering times \cite{Abrikosov98prb,Lu15Weyl-shortrange,Vasilopoulos84jmp}. To sum up, the peak positions follow the relation $\rho_{zz} \sim \sigma_{zz}^{-1} \sim  \sigma_{yy} \sim \rho_{xx} $, thus $\rho_{zz}$ and $\rho_{xx}$ present peaks near Landau band edges and their phase shifts are the same.

\begin{figure}
\centering
\includegraphics[width=0.43\textwidth]{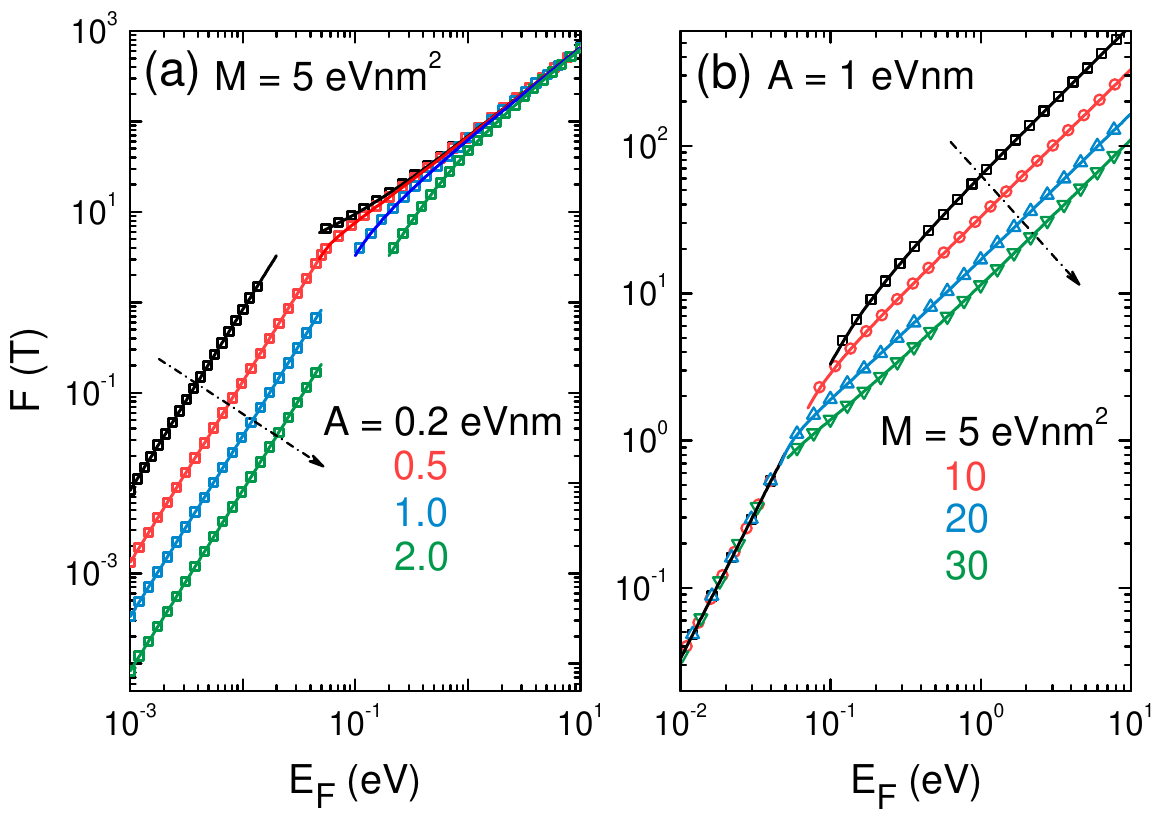}
\includegraphics[width=0.45\textwidth]{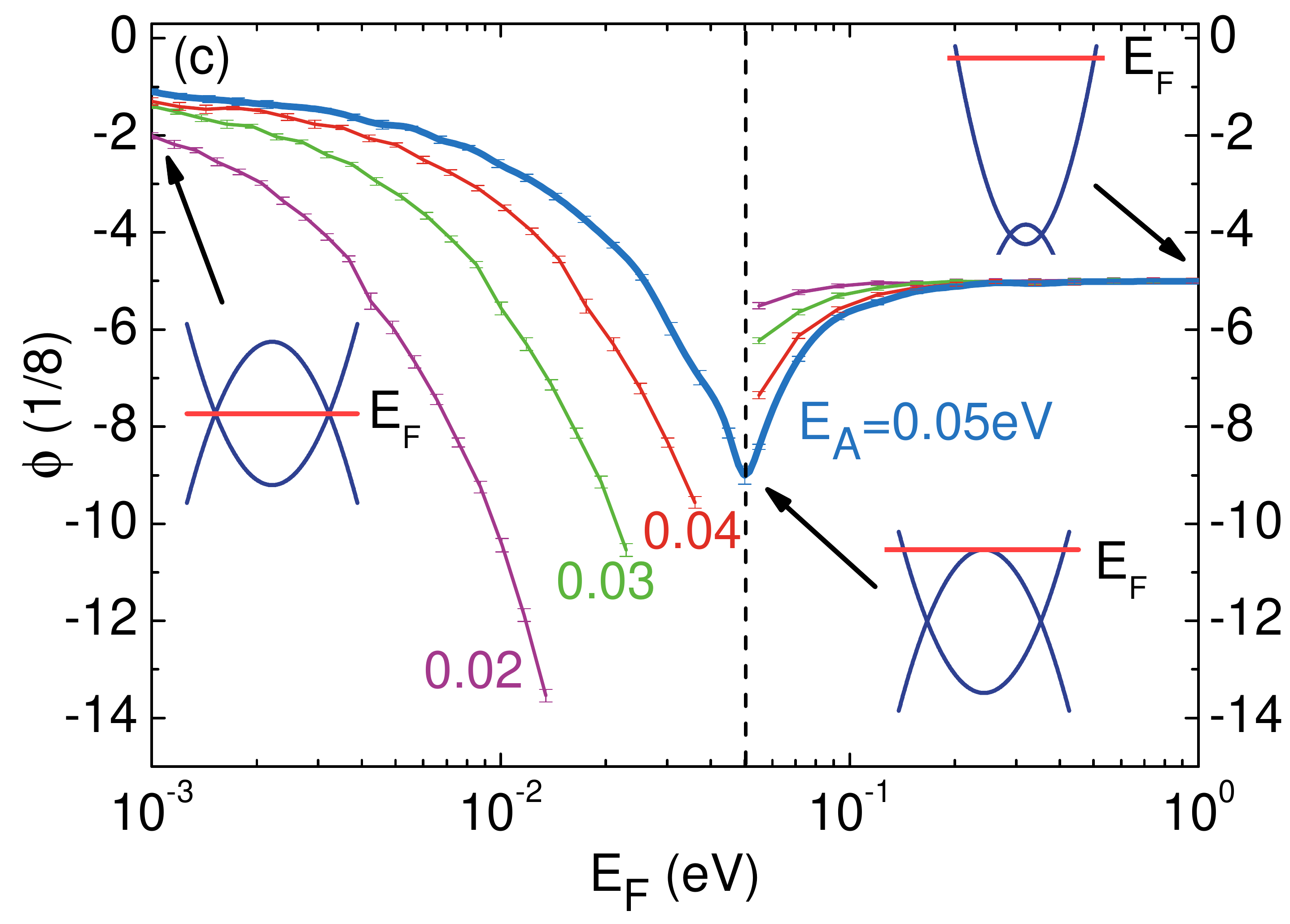}
\caption{Adapted from Ref. \cite{WangCM16prl}. For the Weyl semimetal described by Eq. (\ref{OC-Ham}). (a) The frequency $F$ obtained numerically (scatters) and analytically (solid curves) vs the Fermi energy $E_F$ for (a) different $A$ at a fixed $M$; and (b) for different $M$ at a fixed $A$. (c) The phase shift $\phi$ vs $E_F$ for different $E_A=Ak_w$ and a fixed $E_M=Mk_w^2=0.05$ eV. The curves break because $F$ and $\phi$ cannot be fitted when beating patterns form. The insets indicate the location of Fermi energy. The vertical dashed lines mark the Lifshitz point. $k_w=0.1$ nm$^{-1}$. }\label{fig:phase-shift}
\end{figure}

\subsection{Anomalous phase shift near the Lifshitz point of Weyl and Dirac semimetals}

Figure \ref{fig:phase-shift} shows the results of the frequency $F$ and the phase shift $\phi$ calculated numerically for the model in Eq. (\ref{OC-Ham}). In Fig. \ref{fig:phase-shift}(c), the numerical results are in agreement with the analytical prediction $\phi=-1/8$ for the linear limit ($E_F\rightarrow 0$) and $-5/8$ for the parabolic limit ($E_F\rightarrow \infty $).
We define $E_A= A k_w$ and $E_M=Mk_w^2$.
For $E_M\neq E_A$, the $\phi$-$E_F$ curves break as the beating patterns emerge. In Fig. \ref{fig:phase-shift} (c), for $E_A<E_M$, the phase shift drops below $-5/8$ rather than shift monotonically from -1/8 to -5/8 around the Lifshitz transition point (namely, $E_F=E_M$). This is because there is no simple  $k_z^2$ dependence \cite{WangCM16prl}. At the Lifshitz point, we can analytically show that the phase shift is $-9/8$, which is in agreement with that in Fig.~\ref{fig:phase-shift} (c). It is equivalent to $-1/8$, which is usually considered to originate from the $\pi$ Berry phase as the Fermi sphere encircles the single Weyl node. Nevertheless, the Fermi sphere encircles two Weyl nodes when the Fermi energy is at the Lifshitz point. For $E_A> E_M $, there is no nonmonotonicity in $\phi-E_F$.

A Weyl semimetal combine with its time-reversal counterpart can compose a Dirac semimetal. The model of a Dirac semimetal can be made by $H(\mathbf{k})$ in Eq. (\ref{OC-Ham}), integrated with its time-reversal counterpart $H^*(-\mathbf{k})$, where the asterisk indicates complex conjugate. This model can be treated as a building block for Weyl semimetals, which respect time-reversal symmetry and meanwhile break inversion symmetry \cite{Huang15nc,Weng15prx,Lv15prx,Xu15sci-TaAs,Yang15np,Liu15nmat,ZhangCL16nc,HuangXC15prx, Xu15np-NbAs,Xu16nc-TaP}. Here, for the Dirac semimetal, the total phase shift may take two values, $-1/8$ for $\alpha\in [0,1/4]$ and $[3/4,1]$ or $-5/8$ for $\alpha\in [1/4,3/4]$. Around the Lifshitz point, the total phase shift may change between the two values.

\begin{table}[htbp]
    \centering
\caption{Adapted from Ref. \cite{WangCM16prl}. The phase shift $\phi_{\text{exp}}$ extracted from the experiments on Cd$_3$As$_2$. According to the theory in this work, if peaks from two Weyl components can be distinguished, the phase shift should be $ \phi_{\text{Weyl}}=\phi_{\text{exp}}-1$; otherwise, the phase shift should be $\phi_{\text{Dirac}}=-5/8$. }\label{tab:eps-phase}
\begin{ruledtabular}
\begin{tabular}{cccc}
    Ref.      & $\phi_{\text{exp}}$ &  $\phi_{\text{Weyl}}$ & $\phi_{\text{Dirac}}$\\
 \hline
 \cite{He14prl}   & 0.06 $\sim$ 0.08& -0.94 $\sim$ -0.92 &  -5/8  \\
 \cite{Zhao15prx}  &0.11 $\sim$ 0.38&-0.89 $\sim$ -0.62 &  -5/8\\
\cite{Narayanan15prl}  & 0.04\footnote{Read from Fig 2(d) of Ref. \cite{Narayanan15prl}.} &-0.96 &  -5/8
\end{tabular}
\end{ruledtabular}
\end{table}

The electron carriers is supposed to yield negative phase shifts and the hole carriers is supposed to yield positive phase shifts \cite{Murakawa13sci}. Nevertheless, in experiments of  the Dirac semimetal Cd3As2, the phase shift for electron carriers takes positive values \cite{He14prl,Narayanan15prl,Zhao15prx}. One explanation may be that, for the phase shift $1/8$ to $3/8$ in the experiments, their actual values are around $-7/8$ to $-5/8$ due to the $2\pi$ periodicity. Our numerical results show that the total phase shift adopts these values from around the Lifshitz point to higher Fermi energies, which is also consistent with the carrier density in the experiments. In Table \ref{tab:eps-phase}, we propose the counterparts for the experimental values of the phase shift. In Ref. \cite{WangCM16prl}, we also demonstrate that beating patterns will emerge because of the band inversion, that is different from orbital quantum interference \cite{Xiong16epl}, the Zeeman splitting \cite{ZhangCL15arXiv-TaAs,Hu16srep,Cao15nc} and nested Fermi surfaces \cite{Zhao15prx}.

\section{Strong field: Quantum oscillation in nodal-line semimetals}\label{Sec:QO-NL}

\begin{figure}[tbph]
\centering \includegraphics[width=0.45\textwidth]{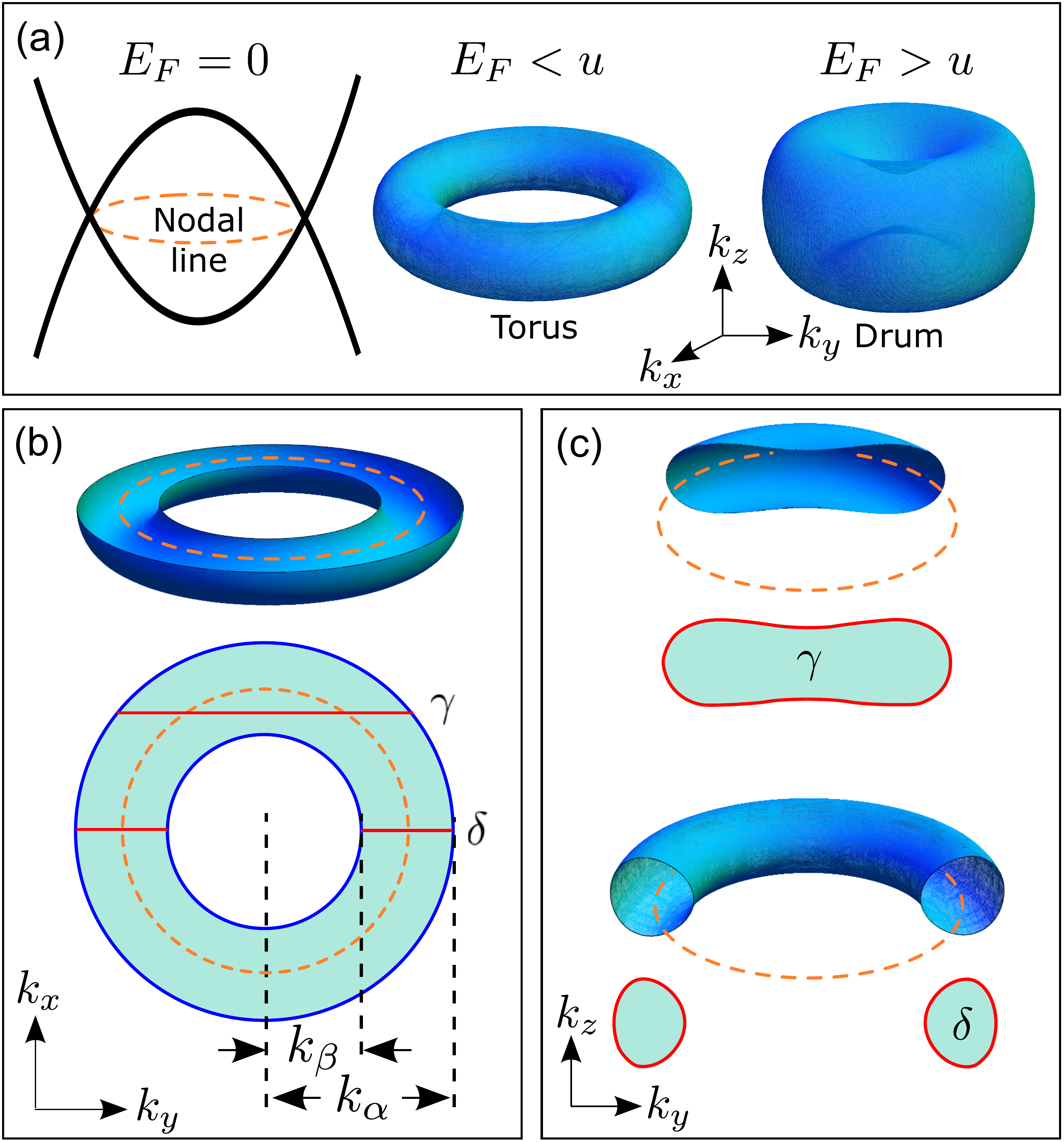}
\caption{Adapted from Ref. \cite{LiC18prl}. (a) The nodal line (dashed ring), torus, and drum Fermi surfaces for a generic model of a nodal-line semimetal in Eq. \eqref{eq:model}. $E_F$ is the Fermi energy. $u$ is a model parameter. (b) The maximum ($\alpha$) and minimum ($\beta$) cross sections in the nodal-line plane of the torus Fermi surface. (c) The maximum ($\gamma$) and minimum ($\delta$) cross sections out of the nodal-line plane of the torus Fermi surface. }
\label{fig:torus}
\end{figure}

\subsection{Phase shifts of nodal-line semimetals}

The frequencies $F$ and phase shifts $\phi$ can be analyzed from the Fermi surface described by Eq. (\ref{eq:NL-model}).
From the Onsager relation, we have $F=(\hbar/2\pi e) A$, where $A$ is the area of the extremal cross section on the Fermi surface perpendicular to the magnetic field \cite{Onsager52pm}. When the nodal-line plane is perpendicular to a magnetic field, specifically $B_z$ here, there are two extremal cross sections at the $k_z=0$ plane [Fig. \ref{fig:torus}(b)] .
Combining this dispersion with the Onsager relation, we find that the high frequency is $F_\alpha=m(u+E_F)/\hbar e$ for the outside circle and the low frequency is $F_\beta=m(u-E_F)/\hbar e$ for the inside circle. These two frequencies may lead to a beating pattern \cite{Phillips14prb}.

\begin{table}[htb]
    \centering
\caption{ Adapted from Ref. \cite{LiC18prl}. The phase shift $\phi$ of the nodal-line semimetal in Fig. \ref{fig:torus}, obtained by using the relation $\phi=-1/2+\phi_{\it B}/2\pi+\phi_{\rm 3D}$, where $\phi_{\it B }$ is the Berry phase and $\phi_{\rm 3D}$ is the dimension correction. $\alpha,\beta,\gamma,\delta$ are the extremal cross sections in Fig. \ref{fig:torus}. Max. or Min. indicates whether the cross section is maximum or minimum. For electron carriers, $\phi_{\rm 3D}$ is $-1/8$ for maximum cross section and $1/8$ for minimum cross section. The phase shifts of hole carriers are opposite to those of electrons.}\label{tab:rule}
\begin{ruledtabular}
\begin{tabular}{ccccc}
     & Berry       & Min.          &  Electron            & Hole      \\
     & phase       &  /max.   &         &      \\ \hline
$\alpha$          &   0          & Max.  & $-1/2+0-1/8=-5/8$ & $+5/8$\\
$\beta$          &   0          & Min.  & $-1/2+0+1/8=-3/8 \leftrightarrow 5/8$ & $ -5/8$\\
$\gamma$          &   0          & Max.  & $-1/2+0-1/8=-5/8$ & $+5/8$\\
$\delta$          &   $\pi$          & Min.  & $-1/2+\pi/2\pi+1/8=1/8$ & $-1/8$\\
\end{tabular}
\end{ruledtabular}
\end{table}

Phase shifts are more complicated in the nodal-line semimetals. First of all, whether the magnetic fields are in-plane or out-of-plane, the torus Fermi surface has both maximum  cross section and minimum cross section. Secondly, the Berry phase is 0  along the circle parallel to the nodal line and is $\pi$ along the circle enclosing the nodal line. Therefore, depending on direction of  the magnetic field, the quantum oscillation allows different phase shifts,  which is summarized in Table \ref{tab:phase}. When the nodal-line plane is perpendicular to a magnetic field, there are two cross sections $\alpha$ and $\beta$, shown in Fig. \ref{fig:torus}. Along loops of the cross sections $\alpha$ and $\beta$, the Berry phase is 0, thus phase shifts take values $ 5/8$ or $-5/8$.
For the electron carriers, (see Table \ref{tab:rule}), the phase shifts of  the maximum cross section$ \alpha$ are $\phi_{\rm 3D}=-1/8$ and $\phi=-1/2+0-1/8=-5/8$.
The phase shifts of the minimum cross section $\beta$ are $\phi_{\rm 3D}=1/8$ and $\phi=-1/2+0+1/8=-3/8$, that is equivalent to $5/8$ since the oscillation has  $2\pi$ periodicity.

\begin{table}[htb]
    \centering
\caption{Adapted from Ref. \cite{LiC18prl}. The analytic expressions for the frequencies $F_{\alpha, \beta}$ and phase shifts $\phi_{\alpha, \beta}$ in the resistivity formula Eq. \eqref{rho} for electron carriers. Hole carriers have an extra minus sign in all cases compared to electron carriers. }\label{tab:freq-pha}
\begin{ruledtabular}
\begin{tabular}{ccccc}
                           & \multicolumn{2}{c}{Longitudinal $\rho_{zz}$} & \multicolumn{2}{c}{Transverse $\rho_{xx}$} \\
                     &$\alpha$                 &$\beta$       &$\alpha$     &$\beta$\\ \hline
\rule{0pt}{0.65cm}$F$      & $\frac{m}{\hbar e}(u+E_F)$       & $\frac{m}{\hbar e}(u-E_F)$           &  $\frac{m}{\hbar e}(u+E_F)$            & $\frac{m}{\hbar e}(u-E_F)$      \\
\rule{0pt}{0.65cm}$\phi$             & $-5/8$           & $+5/8 $ & $- 5/8$ & $+5/8$\\
\end{tabular}
\end{ruledtabular}
\end{table}

\subsection{Magnetoresistivity of nodal-line semimetals}

To testify the above conclusion, we calculate the resistivities along the $z$ ($\rho_{zz}$, out of the nodel-line plane) and $x$ ($\rho_{xx}$, in the nodel-line plane) directions, respectively, according to $\rho_{zz}=1/\sigma_{zz}$ and $\rho_{xx}=\sigma_{yy}/(\sigma_{yy}^2+\sigma_{xy}^2)$. The calculations show that for both $\rho_{xx}$ and $\rho_{zz}$, there are two terms in the magnetoresistivities,
\begin{align}\label{rho}
(\rho-\rho_0)/\rho_0&=\mathcal{C_\alpha}\exp(-\lambda_D)\cos[2\pi(F_\alpha/B+\phi_\alpha)]\nonumber\\
&+\mathcal{C_\beta}\exp(-\lambda_D)\cos[2\pi(F_\beta/B+\phi_\beta)],
\end{align}
where  $\phi_{\alpha, \beta}$ is the phase shifts and $F_{\alpha, \beta}$ are the oscillation frequencies. Their analytic expressions are listed in Table \ref{tab:freq-pha}. We show that the resistivity calculations and the Fermi surface analysis are equivalent for the phase shifts of the quantum oscillation in the nodal-line semimetal. The phase shifts, in a magnetic field parallel to the nodal-line plane [Fig. \ref{fig:torus}(c)], can also be found in a similar way, as listed in Table \ref{tab:rule}.

\subsection{Discussions}

For nodal-line semimetals, most of the quantum oscillation experiments have been done for the ZrSiS family materials \cite{Singha17pnas,Ali16sa,WangX16aem,Lv16apl,Hu16prb,PanH18srep,
Hu16prl,Hu17prb,Kumar17prb}, in which there are both electron and hole pockets at the Fermi energy \cite{Schoop16nc,Neupane16prb,Singha17pnas,PanH18srep}. ZrSiS (see Figure 3 In Ref. \cite{LiC18prl})  has the diamond-shaped electron pockets encircling the nodal line and the quasi-2D tubular-shaped hole pockets at the {\it X} points \cite{PanH18srep}. When the magnetic field is normal to the diamond-shaped Fermi surface, there are three extremal cross sections, the outer ($\alpha$) and inner ($\beta$) cross sections of the diamond-shaped electron pocket, and the tubular-shaped hole pockets ($\gamma$). The $\alpha$ and $\beta$ cross sections of the diamond-shaped Fermi surface take phase shifts of $5/8$ and $-5/8$, respectively. However, the frequencies of the $\alpha$ and $\beta$ pockets are so large (about $10^4$ T) that it is hard to be extracted from the experiments. In contrast, the $\gamma$ nodal-line hole pockets have been identified as the origin of the high frequency (about 210 T) component \cite{PanH18srep}. This pocket encircles a nodal line, thus it has a $\pi$ Berry phase. In addition, it has $\phi_{\rm 3D}=0$ due to its quasi-2D nature. From Eq. (\ref{eq:phase}), we find that the $\gamma$ pocket has a phase shift of $\phi=-1/2+\pi/2\pi+0=0$, which is in agreement with the results obtained by Ali \emph{et al.} \cite{Ali16sa}. In Ref. \cite{LiC18prl}, we also analyzes the phase shifts for Cu$_3$PdN \cite{Yu15prl,Kim15prl}.

When the symmetry protecting the nodal line is broken, there is a finite gap $\Delta$ that separates the conduction and valence bands. The Berry phase becomes
\begin{align}
\phi_{\rm B}=\pm\pi\left(1-\frac{\Delta}{2E_F}\right).
\end{align}

The nodal-line semimetals may also be distinguished from their weak localization behaviors that cross between 2D weak anti-localization and 3D weak localization \cite{ChenW19arXiv}.

\section{Strong field: 3D quantum Hall effect}\label{Sec:3DQHE}

\begin{figure}[htbp]
\centering
\includegraphics[width=0.9\columnwidth]{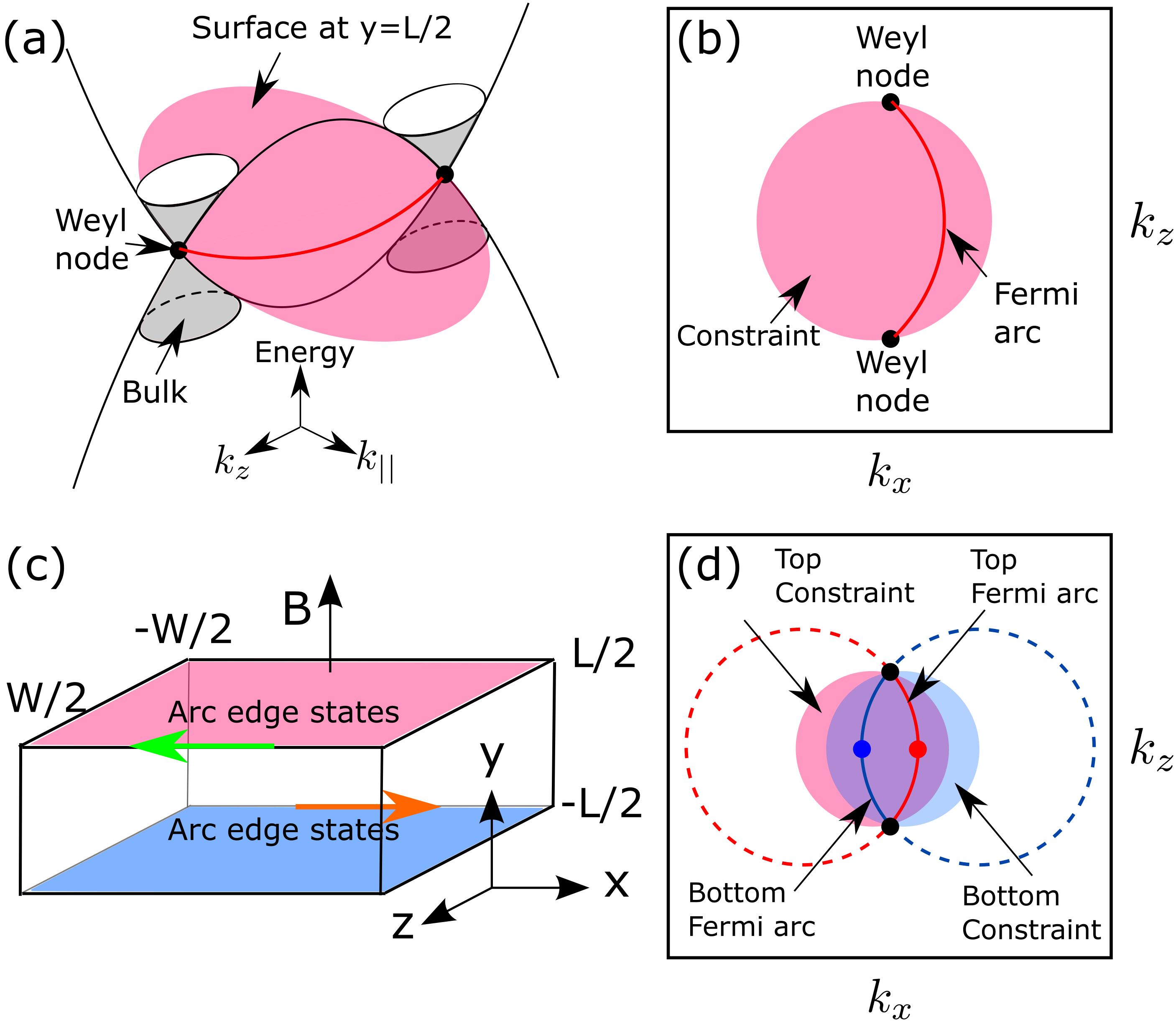}
\includegraphics[width=0.9\columnwidth]{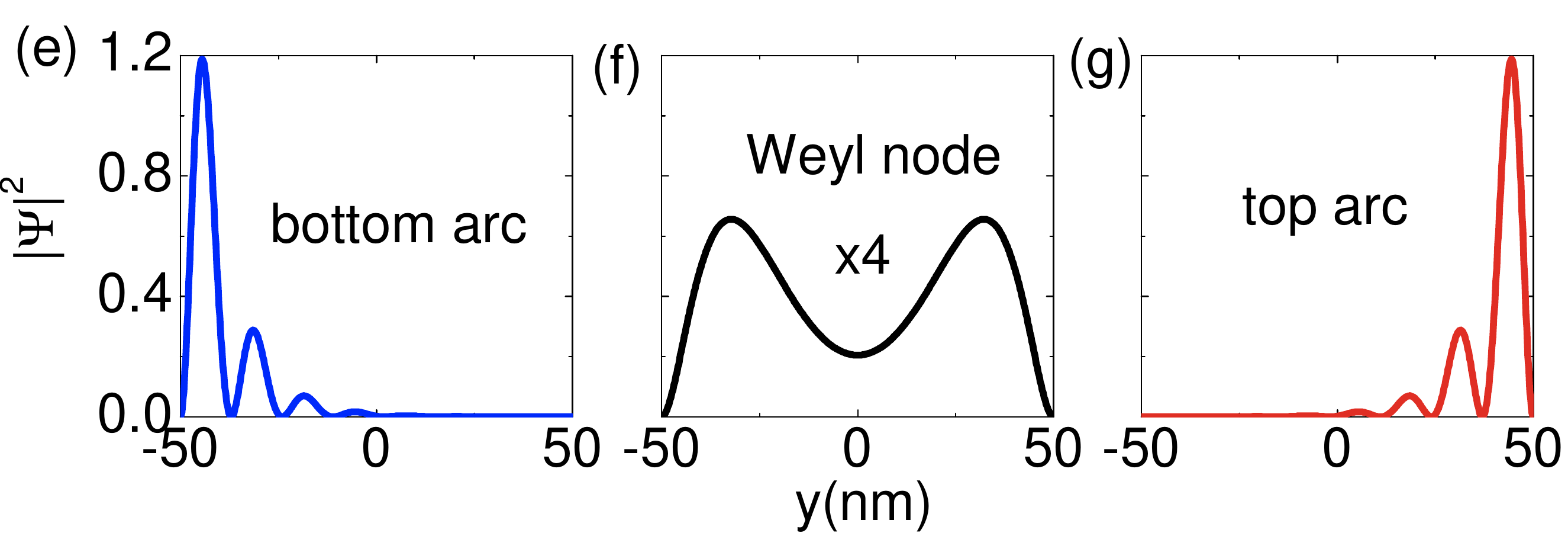}
\includegraphics[width=0.45\columnwidth]{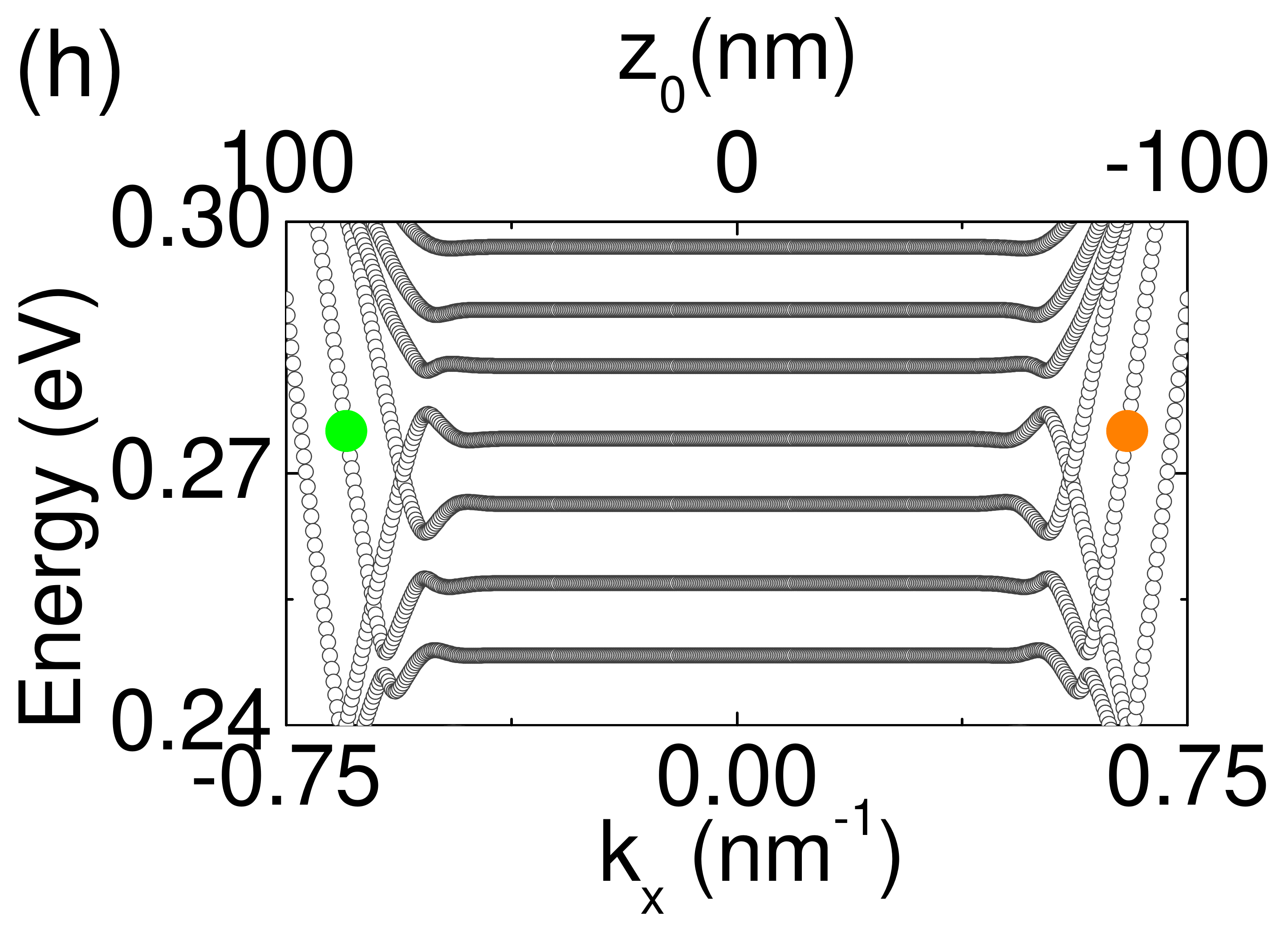}
\includegraphics[width=0.45\columnwidth]{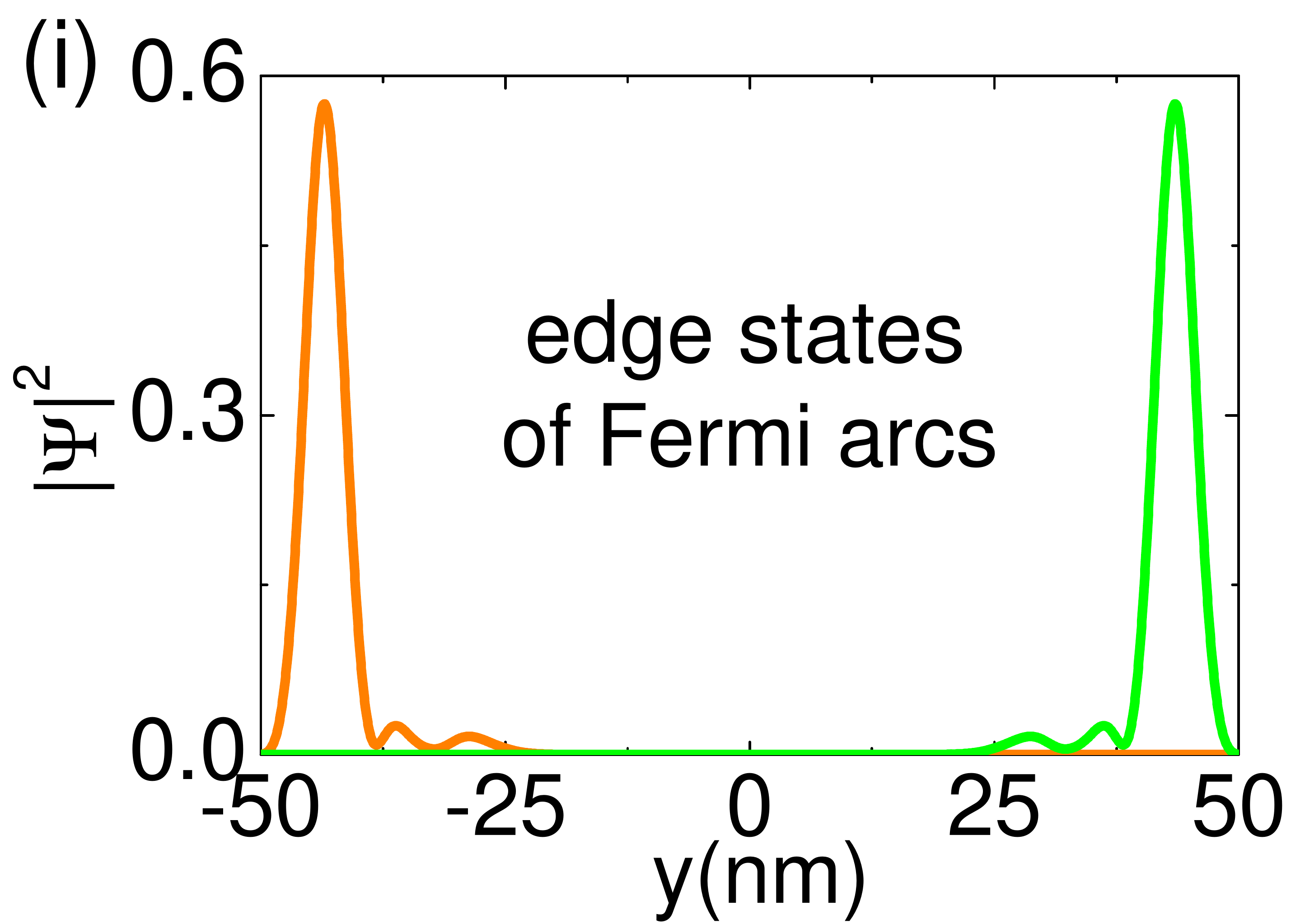}
\caption{Adapted from Ref. \cite{WangCM17prl}. (a) The energy dispersions for the Fermi arc (at $y=L/2$) and bulk states in a topological Weyl semimetal. $k_{||}$ stands for $(k_x,k_y)$ for the bulk and $k_x$ for the arc, respectively. (b) The Fermi arc at $y=L/2$ and $E_F=E_w$ on the $k_z-k_x$ plane. The shadow defines the ``constraint" region where the Fermi arcs can exist. (c) A slab of topological semimetal of thickness $L$ and width $W$. (d) The Fermi arcs at $E_F=E_w$ (solid) and constraints (shadow) at the $y=L/2$ (red) and $-L/2$ (blue) surfaces of the slab. [(e)-(g)] The wave function distributions at $k_z=0$ along the $y$ axis, at the blue (bottom arc, $k_x<0$), black (Weyl nodes), and red (top arc, $k_x>0$) dots in (d). (h) Landau levels of the Fermi arcs at $B=5$ T vs. the guiding center $z_0$. (i) The wave function distributions along the $y$ axis for the edge states of the Fermi arcs marked by the green and orange dots in (h). $L=100$ nm, $W=200$ nm, and other parameters can be found in Fig. \ref{Fig:Hall}. }\label{Fig:arcs}
\end{figure}

The discovery of the quantum Hall effect in 2D opens the door to the field of topological phases of matter \cite{Klitzing80prl,Thouless82prl}.
In 3D electron gases, the extra dimension along the magnetic field direction prevents the quantization of the Hall conductance. Thus, the quantum Hall effect is normally observed in 2D systems \cite{Klitzing80prl,Novoselov05nat,ZhangYB05nat,Xu14np,Yoshimi15nc}.
In Ref. \cite{WangCM17prl}, we show a 3D quantum Hall effect in a topological semimetal. The topological semimetal can be considered as a 2D topological insulator for momenta ($k_z$ here) between the Weyl nodes, resulting in the topologically protected surface states [in Fig. \ref{Fig:arcs} (c)]  at the surfaces parallel to the Weyl node separation direction. The topologically protected states form the Fermi arcs \cite{Brahlek12prl,Wu13natphys,Wang12prb,Liu14sci,Wang13prb,Xu15sci,Wang13prb,Liu14natmat,Neupane14nc,Yi14srep,Borisenko14prl,Weng15prx,Huang15nc,Lv15prx,Xu15sci-TaAs,Liu15nmat} on the Fermi surface [red curves in Figs. \ref{Fig:arcs} (a) and \ref{Fig:arcs}(b)].
The transport signature of the Fermi arcs is an intriguing topic \cite{Hosur12prb,Baum15prx,Gorbar16prb,Ominato16prb,McCormick18prb}.

\subsection{Wormhole tunneling via Fermi arcs and Weyl nodes}

The topological nature requires that only a region between the Weyl nodes can be occupied by the states of Fermi arcs \cite{ZhangSB16njp} [Fig. \ref{Fig:arcs}(b)]. At one surface, a closed Fermi loop, which is essential to  the quantum Hall effect,  cannot  be formed by the Fermi arcs. However, in a topological semimetal slab, the Fermi arcs from opposite surfaces [Fig. \ref{Fig:arcs}(c)] can form the required closed Fermi loop [Fig. \ref{Fig:arcs}(d)]. Thus electrons can tunnel between the Fermi arcs at opposite surfaces via the Weyl nodes [Figs. \ref{Fig:arcs}(e)-\ref{Fig:arcs}(g)]. The Fermi loop formed by the Fermi arcs at opposite surfaces via the Weyl nodes can support a 3D quantum Hall effect.
To be specific, the Weyl nodes act like ``wormholes" that connect the top and bottom surfaces, and an electron can complete the cyclotron motion. Since the Weyl nodes are singularities in momentum, the wormhole tunneling can be infinite in real space, according to the uncertainty principle.
The time scale of the ``wormhole" tunneling is 0 as the Weyl node on the top surface and the Weyl node on the bottom surface are the same one and are described by the one coherence wavefunction.
In experimental materials, the tunneling distance is limited by the mean free path, which can be comparable to or longer than 100 nm in high-mobility topological semimetals \cite{HuangXC15prx,Yang15np,Shekhar15np,ZhangCL16nc,He14prl,Liang15nmat,Zhao15prx,Narayanan15prl,Xiong15sci}, even up to 1 $\mu$m \cite{Moll16nat}, thus the thickness in the calculation is chosen to be 100 nm. The wormhole effect has been addressed in different situations in topological insulators \cite{Rosenberg10prb}.
The quantum Hall effect solely from the Fermi arcs requires the bulk carriers to be depleted by tuning the Fermi energy to the Weyl nodes \cite{Ruan16nc}. Compared to the novel quantum oscillations \cite{Potter14nc,Moll16nat}, the quantum Hall effect of the Fermi arcs contributes a quantized complement to the Fermi arc dominant electronic transports.
The Weyl semimetals TaAs family \cite{Weng15prx,Huang15nc,Lv15prx,Xu15sci-TaAs,Xu2016prl-TaAs,HuangXC15prx,Yang15np,Liu15nmat,Shekhar15np,ZhangCL16nc,Belopolski2016prl,Belopolski2016prb-TaAs,Xu2015sciadv} and the Dirac semimetals Cd$_3$As$_2$ and Na$_3$Bi have extremely high mobilities \cite{He14prl,Liang15nmat,Zhao15prx,Narayanan15prl,Xiong15sci} required by the quantum Hall effect. Low carrier densities \cite{LiCZ15nc,LiH16nc,ZhangC17nc} and gating \cite{LiCZ15nc} have also been achieved. The 3D quantum Hall effect of the Fermi arcs is expected in slabs of the TaAs family \cite{Weng15prx,Huang15nc,Lv15prx,Xu15sci-TaAs,HuangXC15prx,Liu15nmat,Yang15np,Shekhar15np,ZhangCL16nc,Ruan16nc}, [110] or [1$\bar{1}$0] Cd$_3$As$_2$ \cite{Uchida17MarchMeeting,ZhangC17nc,Uchida17nc,Schumann18prl,ZhangC18nat}, and [100] or [010] Na$_3$Bi.

\begin{figure}
\centering
\includegraphics[width=0.49\columnwidth]{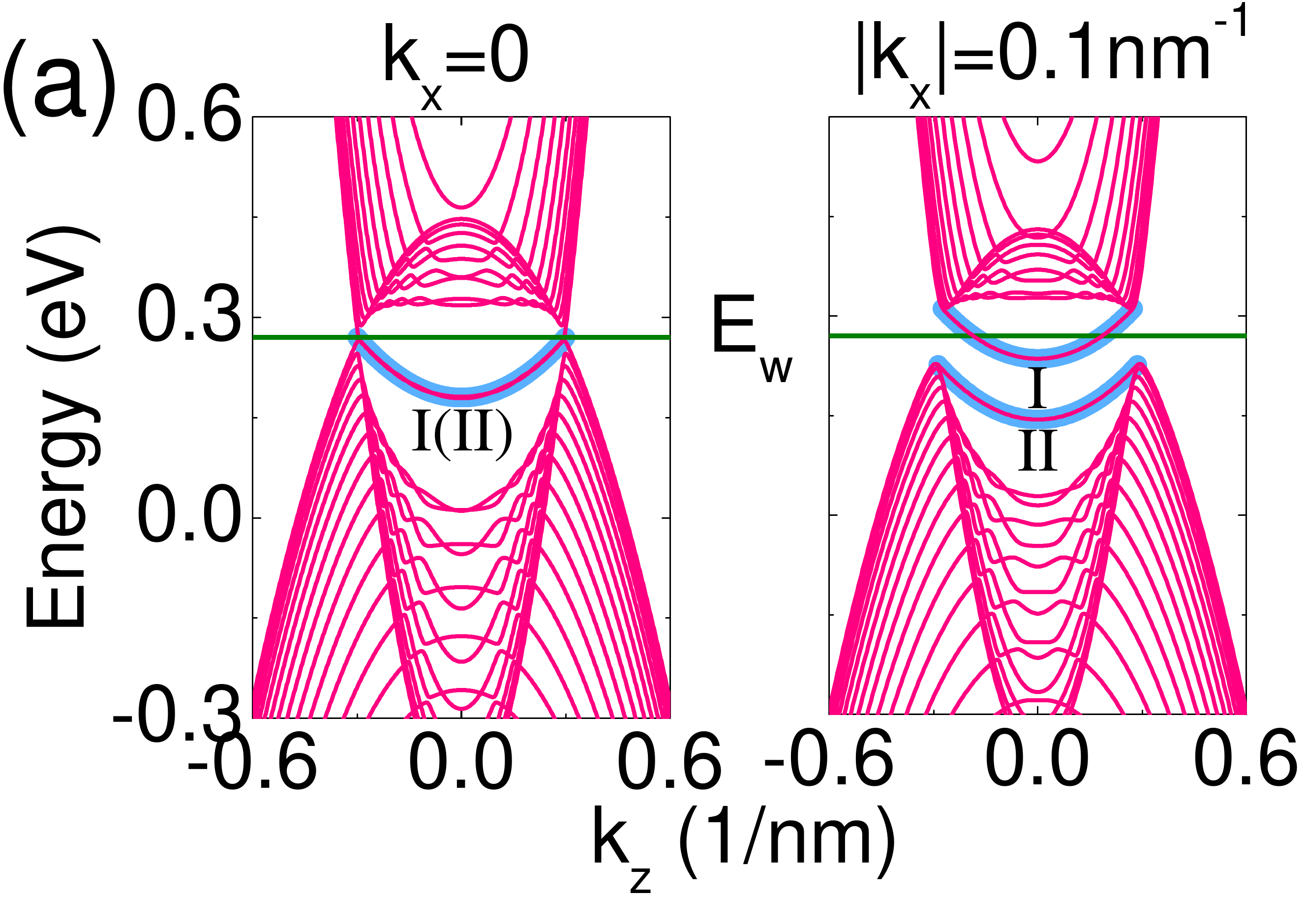}
\includegraphics[width=0.49\columnwidth]{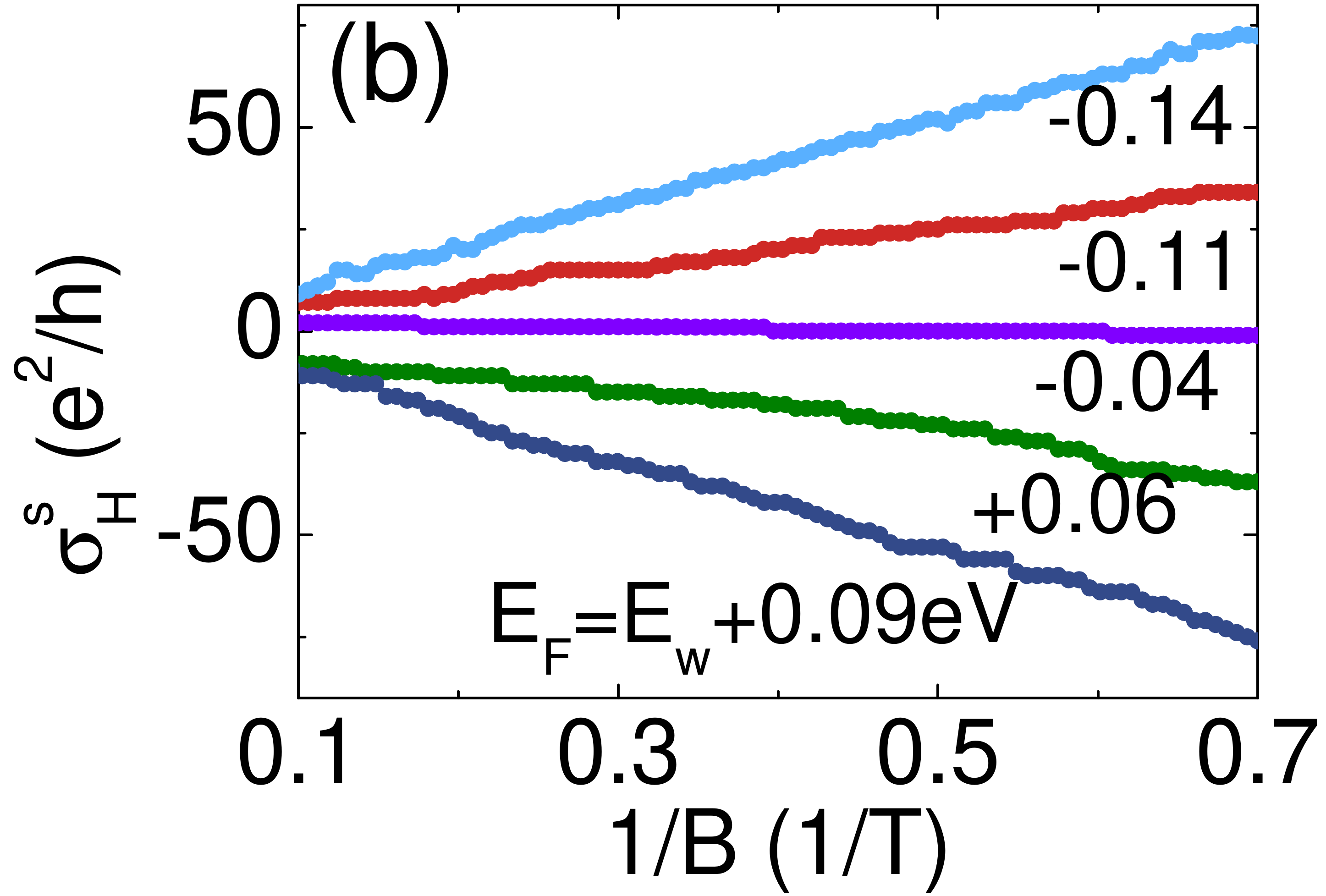}
\includegraphics[width=0.95\columnwidth]{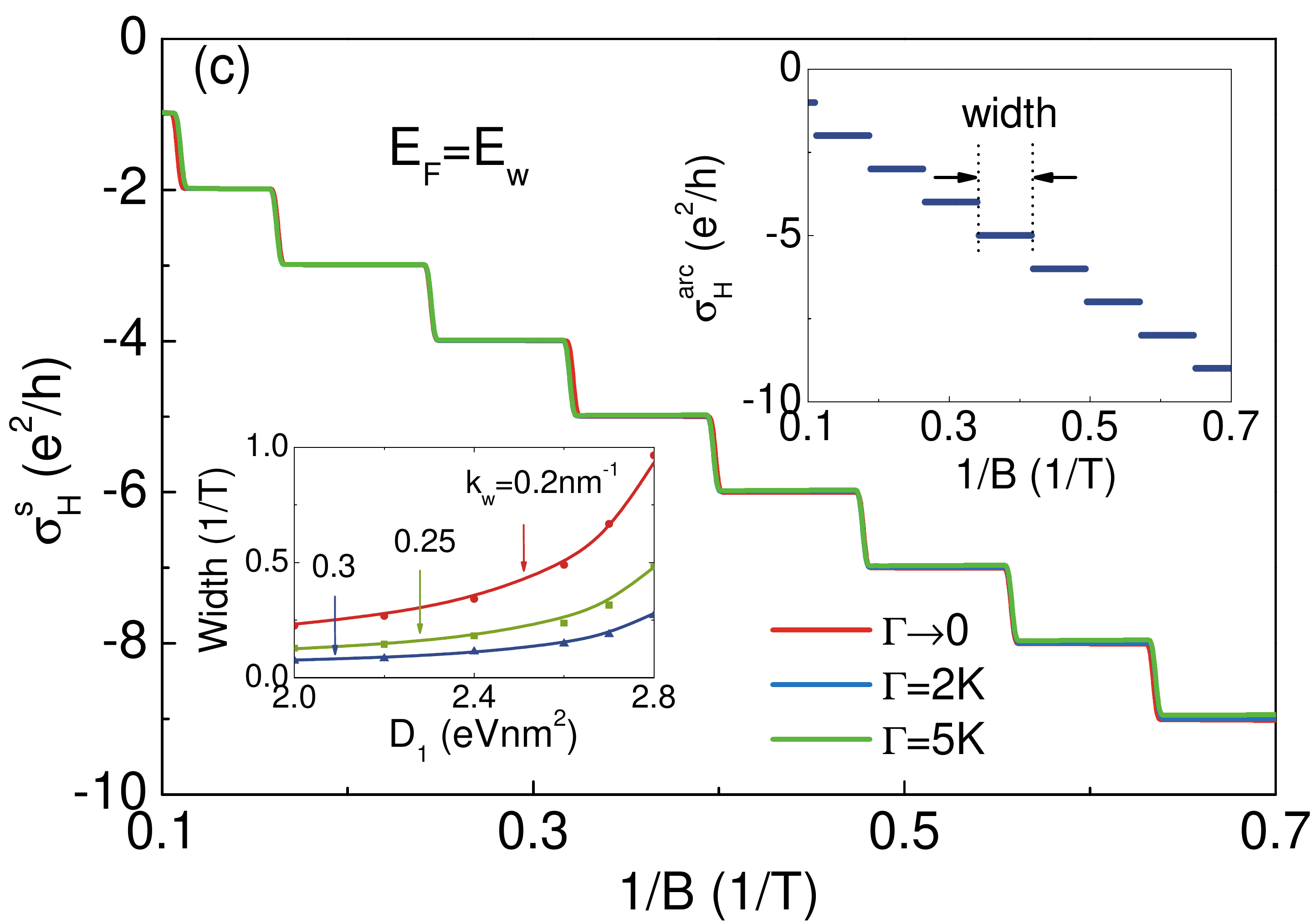}
\caption{ Adapted from Ref. \cite{WangCM17prl}. (a) In a topological semimetal slab, the numerically calculated energy spectrum (pink) for the bulk states and Fermi arcs at $k_x=0$ (left) and $k_x=\pm 0.1$ nm$^{-1}$ (right). The blue curves are the Fermi arc bands plotted using $H_{\textrm{arc}}$ and $H_{\textrm{arc}}'$. (b) The sheet Hall conductivity when the Fermi energy $E_F$ crosses the bulk states for $\Gamma\rightarrow 0$. $\Gamma$ is the disorder-induced level broadening. Recent experiments show that gating can tune carriers from $n$- to $p$-type in 100-nm-thick devices of topological semimetal \cite{LiCZ15nc}. (c) The sheet Hall conductivity $\sigma_{\rm H}^s$ at $E_F=E_w$, where the Fermi energy crosses only arc I. The right inset shows the analytic Hall conductance $\sigma_{\rm H}$. In the presence of a residual detuning from the Weyl nodes, the bulk states also contribute to $\sigma_{\rm H}^s$. Unlike that from the Fermi arcs, the contribution from the bulk states may change with the slab thickness. The left inset shows the width of the Hall plateaus in the clean limit as a function of $D_1$ for different $k_w$. The dots and lines are the numerical and analytic results, respectively. The parameters are $M$=$5$eVnm$^2$, $A$=0.5eVnm, and $D_2$=$3$eVnm$^2$, $D_1$=$2$eVnm$^2$, $k_w$=$0.3$nm$^{-1}$, and $L$=$100$ nm.}\label{Fig:Hall}
\end{figure}

\subsection{Quantized Hall conductance}

We can calculate the Hall conductivity from the Kubo formula \cite{Thouless82prl,Gusynin05prl,Zyuzin11prb,ZhangSB14prb,ZhangSB15srep,Pertsova16prb,WangCM17prl}.
Figure \ref{Fig:Hall} (b) presents the sheet Hall conductivity  $\sigma_{\rm H}^s$ for the topological semimetal slab. When the Fermi energy is far away from the Weyl nodes, the sheet Hall conductivity obeys the usual $1/B$ dependence.
The closer the Fermi energy moves towards the Weyl nodes, the smaller the slope becomes, which indicates that the carrier density is decreasing. Moreover, when the Fermi energy moves  towards the Weyl nodes, the quantized plateaus of $\sigma_{\rm H}^s$ begin to arise. Note that a 100-nm slab is still a 3D object.

\subsection{3D distribution of the edge states}

Figures \ref{Fig:arcs}(h) and \ref{Fig:arcs}(i) show that the edge states of the Fermi arcs have a unique 3D  spatial distribution. Specifically, the top edge states propagate to the left (green arrow) and the bottom edge states to the right (orange arrow). This unique 3D distribution of the edge states of the Fermi arcs can be probed by scanning tunneling microscopy \cite{ZhengH16acsnano} or microwave impedance microscopy \cite{Ma15nc}. Different from topological insulators \cite{Xu14np,Yoshimi15nc}, the Fermi-arc quantum Hall effect requires the collaboration of the two surfaces.

\subsection{Topological Dirac semimetals}

A single surface of the Dirac semimetal can form a complete Fermi loop needed by the quantum Hall effect due to the time-reversal symmetry. However, the single surface Fermi arc loop is not that robust and may be distorted \cite{Kargarian16pnas}. Therefore, it may present different characteristics compared to the two-surface Fermi arc loop. For the [112] and [110] Cd$_3$As$_2$ and [010] Na$_3$Bi \cite{Xu15sci} slabs, the parameters from Ref. \cite{Cano17prbrc} yield
the 3D quantum Hall effect, which may exhibit a fourfold degeneracy.

\section{Extremely strong field: Weyl fermion annihilation}\label{Sec:Annihilation}

The topological properties of the Weyl nodes can be revealed by studying the high-field transport properties of a Weyl semimetal. The lowest Landau bands of the Weyl cones remain at zero energy unless a strong magnetic field couples the Weyl fermions of opposite chirality. The coupled Weyl fermions lost their chiralities and acquire masses, two of the most characteristic features of the Weyl fermion. In this sense, the Weyl fermions are annihilated. In the Weyl semimetal TaP, we achieve such a magnetic coupling \cite{ZhangCL17np}. Their lowest Landau bands move above chemical potential, leading to a sharp sign reversal in the Hall resistivity at a specific magnetic field corresponding to the W$_1$ Weyl node separation.
In the following, we use a model calculation to show the physics.

\subsection{Calculation of Landau bands with magnetic fields in the x-z plane}

In Sec. 2.1 of \cite{Lu17fop} and Eq. (\ref{OC-Ham}), we have given
a minimal model for a Weyl semimetal
\begin{eqnarray}\label{eq:model}
H=A(k_{x}\sigma_{x}+k_{y}\sigma_{y})+\mathcal{M_{\mathbf{k}}}\sigma_{z},
\end{eqnarray}
where $\sigma$ are the Pauli matrices, $\mathcal{M}_{\mathbf{k}}=M_{0}-M_{1}(k_{x}^{2}+k_{y}^{2}+k_{z}^{2})$,
$\mathbf{k}=(k_{x},k_{y},k_{z})$ is the wave vector, and $A$, $M_{0/1}$ are
model parameters. When $M_{0}M_{1}>0$, the intersections of the two bands are at $(0,0,\pm k_w)$
where $k_w\equiv\sqrt{M_{0}/M_{1}}$ (see Fig. 1 of \cite{Lu17fop}),  leading to the topological semimetal phase.

In Sec. 2.6 of \cite{Lu17fop}, we have given the Landau bands in a magnetic field along the $z$ direction. Now we generalize the case to an arbitrary field applied normal to the $y$ direction $\mathbf{B} =B(\sin\phi,0,\cos\phi)$, where $\phi$ is the angle between the $z$ and field directions.
The Landau gauge can be chosen as $\mathbf{A}=(-B_z ,0,B_x )y$. Under the Pierls replacement
\begin{eqnarray}
\mathbf{k} \rightarrow  (k_x-y\cos\phi/\ell_B^2,-i\partial_y,k_z+ y\sin\phi/\ell_B),
\end{eqnarray}
the Hamiltonian becomes $h(\mathbf{k}) \rightarrow$
\begin{eqnarray}
\left[
  \begin{array}{cc}j
    \mathcal{M}_k^B & A(k_x-y\cos\phi/\ell_B^2 -\partial_y) \\
    A(k_x-y\cos\phi/\ell_B^2 +\partial_y) & -\mathcal{M}_k^B \\
  \end{array}
\right],\nonumber\\
\end{eqnarray}
where $\ell_B^2 \equiv \hbar/e|B| $ and
$ \mathcal{M}_k^B = M_1[k_c^2-(k_x-y\cos\phi/\ell_B^2 )^2+\partial_y^2-(k_z+y\sin\phi/\ell_B^2)^2]$.
Define the guiding center
\begin{eqnarray}
y_0 &=&  \ell_B^2 (k_x\cos\phi -k_z\sin\phi),
\end{eqnarray}
and the ladder operators \cite{Shen04prl}
\begin{eqnarray}
a = - \frac{1}{\sqrt{2}}(\frac{y-y_0}{\ell_B}  +\ell_B \partial_y ), \ \
a^\dag = - \frac{1}{\sqrt{2}}(\frac{y-y_0}{\ell_B}  -\ell_B \partial_y ),\nonumber\\
\end{eqnarray}
the Hamiltonian becomes
\begin{eqnarray}\label{h-a}
h_a =
\left[
  \begin{array}{cc}
   \mathcal{M}_a & \mathcal{A}^-_a\\
    \mathcal{A}^+_a & -\mathcal{M}_a \\
  \end{array}
\right],
\end{eqnarray}
where
\begin{eqnarray}
\mathcal{M}_a&=&  M_1[k_c^2-k_{||}^2-\frac{2}{\ell_B^2}(a^\dag a+\frac{1}{2})],\nonumber\\
\mathcal{A}^\pm_a
&=&A(k_{||}\sin\phi +\frac{\cos\phi\mp 1}{\sqrt{2}\ell_B}
a + \frac{\cos\phi\pm 1}{\sqrt{2}\ell_B}a^\dag).
\end{eqnarray}
We have defined $k_{||}=k_x\sin\phi +k_z\cos\phi$, which is the summation of the projections of $k_x$ and $k_z$ along the direction of the magnetic field and can serve as a good quantum number.

\subsection{Landau bands in the $z$-direction magnetic field}\label{Sec:Landau-Bz}

At $\phi=0$, i.e., the magnetic field is applied along the $z$ direction, the Hamiltonian reduces to
\begin{eqnarray}
h_a =
\left[
  \begin{array}{cc}
   \mathcal{M}_a & \eta a\\
    \eta a^\dag & -\mathcal{M}_a \\
  \end{array}
\right],
\end{eqnarray}
where $ \mathcal{M}_a=M_0-M_1k_z^2 -\omega (a^\dag a +1/2)$, $\eta=\sqrt{2}A/\ell_B$, $\omega=  2M_1/\ell_B^2$.
With
the trial wave functions $(c_{1}|\nu-1\rangle,c_{2}|\nu\rangle)^{T}$
for $\nu=1,2,...$ (later denoted as $\nu\ge1$) and $(0,|0\rangle)^{T}$
for $\nu=0$, where $\nu$ indexes the Hermite polynomials, the eigen
energies $E$ can be found from the secular equation
\begin{equation}
\det\left[\begin{array}{cc}
\mathcal{M}_{\nu}+\omega/2-E & \eta\sqrt{\nu}\\
\eta\sqrt{\nu} & -\mathcal{M}_{\nu}+\omega/2-E
\end{array}\right]=0
\end{equation}
for $\nu\geq1$, and $-\mathcal{M}_{\nu}+\omega/2-E=0$ for $\nu=0$,
where $\mathcal{M}_{\nu}=M_{0}-M_{1}k_{z}^{2}-\omega\nu$. The eigen
energies are found as
\begin{eqnarray}\label{Energy-LL}
E_{k_{z}}^{\nu\pm} & = & \omega/2\pm\sqrt{\mathcal{M}_{\nu}^{2}+\nu\eta^{2}},\ \nu\ge1\nonumber \\
E_{k_{z}}^{0} & = & \omega/2-M_{0}+M_{1}k_{z}^{2},\ \ \nu=0.
\end{eqnarray}
They represent a set of Landau energy bands ($\nu$ as band index)
dispersing with $k_{z}$. The
eigen states for $\nu\ge1$ are
\begin{eqnarray}\label{LL-nu}
|\nu\ge1,k_{x},k_{z},+\rangle & = & \left[\begin{array}{cc}
\cos\frac{\theta_{k_{z}}^{\nu}}{2}|\nu-1\rangle\\
\sin\frac{\theta_{k_{z}}^{\nu}}{2}|\nu\rangle
\end{array}\right]|k_{x},k_{z}\rangle,\nonumber \\
|\nu\ge1,k_{x},k_{z},-\rangle & = & \left[\begin{array}{cc}
\sin\frac{\theta_{k_{z}}^{\nu}}{2}|\nu-1\rangle\\
-\cos\frac{\theta_{k_{z}}^{\nu}}{2}|\nu\rangle
\end{array}\right]|k_{x},k_{z}\rangle,
\end{eqnarray}
and for $\nu=0$ is
\begin{equation}\label{LL-nu0}
|\nu=0,k_{x},k_{z}\rangle=\left[\begin{array}{cc}
0\\
|0\rangle
\end{array}\right]|k_{x},k_{z}\rangle,\nonumber\\
\end{equation}
where $\cos\theta=\mathcal{M}_{\nu}/\sqrt{\mathcal{M}_{\nu}^{2}+\nu\eta^{2}}$,
and the wave functions $\psi_{\nu,k_{z},k_{x}}(\mathbf{r})=\langle\mathbf{r}|\nu,k_{x},k_{z}\rangle$
are found as
\begin{eqnarray}\label{psi-nukxkz}
\psi_{\nu,k_{z},k_{x}}(\mathbf{r}) & = & \frac{C_{\nu}}{\sqrt{L_{x}L_{z}\ell_{B}}}e^{ik_{z}z}e^{ik_{x}x}e^{-\frac{(y-y_{0})^{2}}{2\ell_{B}^{2}}}\mathcal{H}_{\nu}(\frac{y-y_{0}}{\ell_{B}}),\notag\\
\end{eqnarray}
where $C_{\nu}\equiv1/\sqrt{\nu!2^{\nu}\sqrt{\pi}}$, $L_{x}L_{z}$
is the area of the sample, the guiding center $y_{0}=k_{x}\ell_{B}^{2}$,
$\mathcal{H}_{\nu}$ are the Hermite polynomials. As the dispersions
are not explicit functions of $k_{x}$, the number of different $k_{x}$
represents the Landau degeneracy $N_{L}=1/2\pi\ell_{B}^{2}=eB/h$
in a unit area in the $x-y$ plane.

\subsection{Landau bands in the $x$-direction magnetic field}
\begin{figure}
\centering
\includegraphics[width=0.48\textwidth]{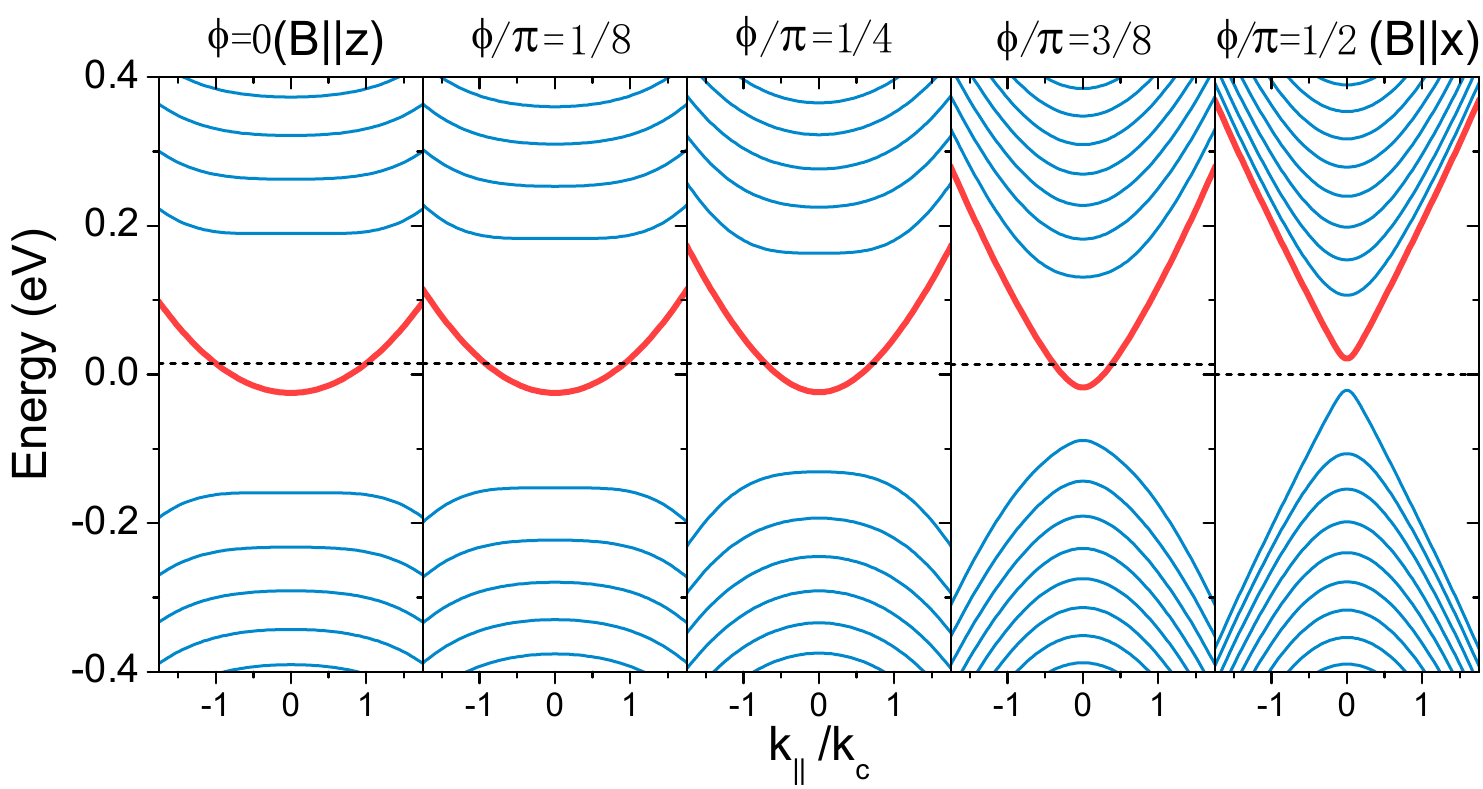}
\caption{For the two-node model for Weyl semimetals (Eq. \ref{eq:model}), the energy spectrum of the Landau bands in a magnetic field $
\mathbf{B}$ applied normal to the $y$ direction, as functions of the wave vector $k_{||}\equiv k_x\sin\phi+k_z\cos\phi$ that is parallel to $\mathbf{B}$. $\phi$ is defined by $\tan\phi=B_x/B_z$. The red curve is the $0$th Landau bands. The dashed line is the Fermi energy. The parameters $k_w=0.2$ nm$^{-1}$, $M_1=1$ eV$\cdot$nm$^2$, $A=1$ eV$\cdot$nm, and $B=10$ T. The number of the Landau bands used in the numerical diagonalization is 401.}
\label{Fig:LB-BzBx}
\end{figure}
If $\phi\neq 0$, the Hamiltonian can be solved numerically. We can write Eq. (\ref{h-a}) as
\begin{eqnarray}
h_a &=& h_a^{(0)} +  h_a',\nonumber\\
h_a^{(0)}&=&
\left[
  \begin{array}{cc}
   \mathcal{M}_a & Ak_{||}\sin\phi\\
    Ak_{||}\sin\phi & -\mathcal{M}_a \\
  \end{array}
\right],\nonumber\\
h_a'&=&
\left[
  \begin{array}{cc}
   0 & \eta(\frac{\cos\phi + 1}{2}
a + \frac{\cos\phi - 1}{2}a^\dag)\\
 \eta (\frac{\cos\phi- 1}{2}
a + \frac{\cos\phi+ 1}{2}a^\dag) & 0 \\
  \end{array}
\right],\nonumber\\
\end{eqnarray}
where $k_{||}=k_x\sin\phi+k_z\cos\phi$,
$\mathcal{M}_a =   M_1[k_c^2-k_{||}^2-\frac{2}{\ell_B^2}(a^\dag a+\frac{1}{2})]$ and $\eta=\sqrt{2}A/\ell_B$.
$H_a^{(0)}$ can be readily diagonalized to give the band spectrum
\begin{eqnarray}\label{E_nu+-_kx}
E^{\nu\pm}_{k} &=& \pm \sqrt{\mathcal{M}_\nu^2+(A k_{||}\sin\phi)^2},
\end{eqnarray}
where $\mathcal{M}_\nu \equiv M_1 [k_c^2- k_{||}^2 - \frac{2}{\ell_B^2}
(\nu +\frac{1}{2})]$.
The wave functions for the conduction and valence bands are
\begin{eqnarray}
|\nu,+,k_{||}\rangle^{(0)} = \left[
  \begin{array}{cc}
    \cos\frac{\Theta^\nu_{k_{||}}}{2}  \\
   \sin\frac{\Theta^\nu_{k_{||}}}{2}   \\
  \end{array}
\right],
\
|\nu,-,k_{||}\rangle^{(0)} = \left[
  \begin{array}{cc}
   \sin\frac{\Theta^\nu_{k_{||}}}{2}  \\
  -\cos\frac{\Theta^\nu_{k_{||}}}{2} \\
  \end{array}
\right]
,\nonumber\\
\end{eqnarray}
with $\cos\Theta^\nu_{k_{||}}\equiv \mathcal{M}_\nu/E^{\nu+}_{k_{||}}$, $\sin\Theta^\nu_{k_{||}}\equiv Ak_{||}\sin\theta /E^{\nu+}_{ k_x}$.
Denote
\begin{eqnarray}
A_{\mu,\nu} &=& A(\frac{\cos\phi+1}{\sqrt{2}\ell_B}\sqrt{\nu} \delta_{\mu,\nu-1} + \frac{\cos\phi-1}{\sqrt{2}\ell_B} \sqrt{\nu+1} \delta_{\mu,\nu+1} ),\nonumber\\
A_{\mu,\nu}^\dag &=& A(\frac{\cos\phi-1}{\sqrt{2}\ell_B}\sqrt{\nu} \delta_{\mu,\nu-1} + \frac{\cos\phi+1}{\sqrt{2}\ell_B} \sqrt{\nu+1} \delta_{\mu,\nu+1} ),\nonumber\\
\end{eqnarray}
we can calculate the off-diagonal matrix elements in the basis of $|\nu,\pm,k_{||}\rangle^{(0)}$ as
\begin{eqnarray}
&&\langle \mu,+,k_{||} |H_a'
|\nu,+,k_{||}\rangle\nonumber\\
&=& A_{\mu,\nu}\cos\frac{\Theta^\mu_{k_{||}}}{2} \sin\frac{\Theta^\nu_{k_{||}}}{2}+ A_{\mu,\nu}^\dag\sin\frac{\Theta^\mu_{k_{||}}}{2} \cos\frac{\Theta^\nu_{k_{||}}}{2},\nonumber\\
&&\langle \mu,-,k_{||} |H_a'
|\nu,-,k_{||}\rangle \nonumber\\
&=& -A_{\mu,\nu}\sin\frac{\Theta^\mu_{k_{||}}}{2} \cos\frac{\Theta^\nu_{k_{||}}}{2} - A_{\mu,\nu}^\dag\cos\frac{\Theta^\mu_{k_{||}}}{2}\sin\frac{\Theta^\nu_{k_{||}}}{2},\nonumber\\
&&\langle \mu,+,k_{||} |H_a'
|\nu,-,k_{||}\rangle \nonumber\\
&=& -A_{\mu,\nu}\cos\frac{\Theta^\mu_{k_{||}}}{2}\cos\frac{\Theta^\nu_{k_{||}}}{2}+ A_{\mu,\nu}^\dag\sin\frac{\Theta^\mu_{k_{||}}}{2}\sin\frac{\Theta^\nu_{k_{||}}}{2},\nonumber\\
&&\langle \mu,-,k_{||} |H_a'
|\nu,+,k_{||}\rangle \nonumber\\
&=& A_{\mu,\nu}\sin\frac{\Theta^\mu_{k_{||}}}{2}\sin\frac{\Theta^\nu_{k_{||}}}{2}- A_{\mu,\nu}^\dag  \cos\frac{\Theta^\mu_{k_{||}}}{2}\cos\frac{\Theta^\nu_{k_{||}}}{2}.\nonumber\\
\end{eqnarray}
Then the energy spectrum along arbitrary directions in the $x-z$ plane can be solved numerically.

Figure \ref{Fig:LB-BzBx} shows the Landau bands of the Weyl semimetal. When the magnetic field is along the $z$ direction ($\theta=0$), the lowest Landau band (red) crosses the Fermi energy (dashed line). As the magnetic field is rotated from the $z$ direction to the $x$ direction ($\theta=\pi/2$), the lowest Landau band is shifted and evolving. When the magnetic field is along the $x$ direction, the spectrum of the Landau bands is particle-hole symmetric and there is a gap due to the coupling between the Weyl fermions at the opposite Weyl nodes. This gap is why there is a sharp sign reversal in the Hall resistivity in the strong-field quantum limit of the Weyl semimetal TaP \cite{ZhangCL17np}. Because of the gap, the Weyl fermions acquire masses and lose their chiralities. Since having chirality and no mass are two features of the Weyl fermion, the Hall signal therefore indicates that the Weyl fermions are annihilated.

\section{Extremely strong field: Forbidden backscattering and resistance dip in the quantum limit}\label{Sec:TI-QL}

The discovery of  3D topological insulators \cite{Hasan10rmp,Qi11rmp,Shen17book}, whose characteristics is of  topologically protected 2D surface states, shines light on the exploring of  exotic topological phases \cite{Yu10sci,Chang13sci,Fu08prl,Akhmerov09prl,Belopolski2017sciadv,Chiu2018epl}. Therefore, distinguishing the bulk-state transport to identify topological insulators is an intriguring topic.

\begin{figure}[htbp]
\centering
\includegraphics[width=0.45\textwidth]{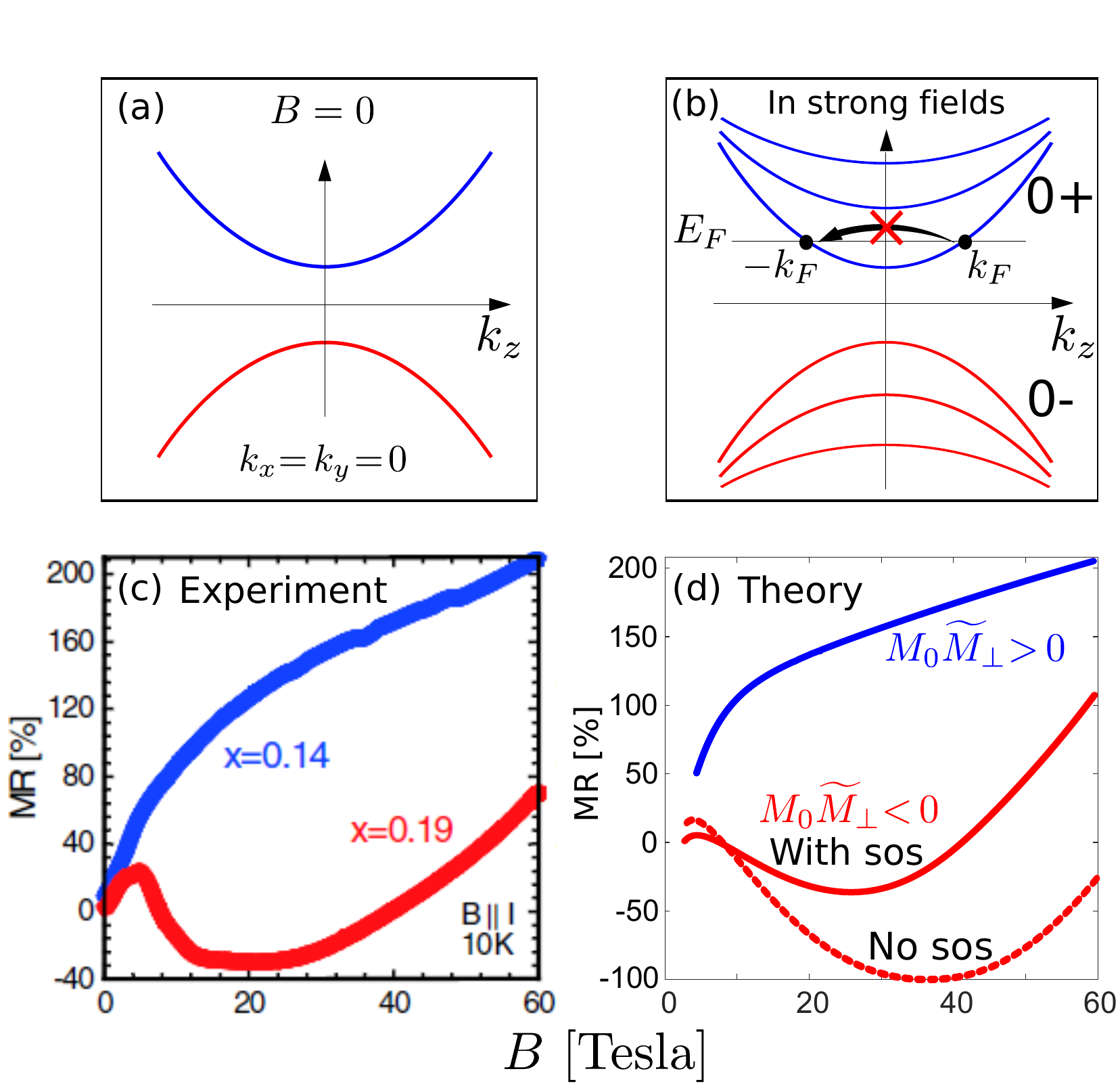}
\caption{ Adapted from Ref. \cite{ChenYY18prl}. In the quantum limit of a 3D topological insulator, the backscattering between the only two states at the Fermi energy can be forbidden at a critical magnetic field, leading to a resistance dip. (a) The zero-field energy spectrum vs $k_z$ of a 3D topological insulator at $k_x=k_y=0$. (b) In a strong magnetic field, the lowest Landau energy bands of the 3D topological insulator vs $k_z$. The Fermi energy $E_F$ crosses only the $0+$ Landau band. $k_F$ and $-k_F$ stand for the only two states at the Fermi energy. (c) The magnetoresistance of Pb$_{1-x}$Sn$_x$Se adapted from Ref. \cite{Assaf17prl}. (d) The calculated magnetoresistance. The abbreviation ``sos" means spin-orbit scattering. The parameters are $M_0=-0.01$ eV, $M_z=0$, $\widetilde{M}_\bot$=18 eV\AA$^2$, $\alpha_1=100$ eVT, and  $\alpha_2=0.0025$ eVT$^{-2}$.}\label{Fig:backscattering}
\end{figure}

In strong magnetic fields, 1D Landau bands are formed from the quantization of the bulk states of a 3D topological insulator. In 2D, lowest Landau levels cross each other, which servers as a signature for the quantum spin Hall phase \cite{Konig07sci,Buttner11np}. In 2D, other approaches can be deployed to probe the quantum spin Hall phase, for example, interference effects \cite{Mani17srep}. However, in 3D, it is seldom addressed that whether the lowest Landau band could be used to identify a topological insulator. In Ref. \cite{ChenYY18prl}, we study the resistance of a 3D topological insulator in the strong-field quantum limit, namely, only the lowest Landau band is occupied [Fig. \ref{Fig:backscattering}(b)]. We find that the backscattering can be totally suppressed in the quantum limit at a critical magnetic field, which can be used to identify the topological insulator phases. Besides, this forbidden backscattering is absent in topological semimetals \cite{Lu15prb-QL,Goswami15prb,ZhangSB16njp}. This theory is consistent with the recent experiment [Figs. \ref{Fig:backscattering}(c) and \ref{Fig:backscattering} (d)].
Moreover, this mechanism will be practical for those materials with small gap, for example, the ZrTe$_5$ \cite{Weng14prx,Liu16nc} families and the Ag$_2$Te \cite{Zhang11prl}.

\subsection{\label{sec:level2}Forbidden backscattering in the quantum limit}

In a strong magnetic field $B$ along the $z$ direction, the energy spectrum quantizes into a series of 1D Landau bands [Figs. \ref{Fig:backscattering}(a) and \ref{Fig:backscattering} (b)].
The energies of the lowest two Landau bands, denoted as $0+$ and $0-$, are $E_{0\pm}=C_0+C_zk_z^2+C_\bot/\ell_B^2\pm\sqrt{m^2+V_n^2k_z^2}$,
where the magnetic length $\ell_B\equiv \sqrt{\hbar/eB}$, the electron charge $-e$, and the mass term
\begin{eqnarray}\label{Eq:m}
m=M_0+M_zk_z^2+ M_\bot/\ell_B^2.
\end{eqnarray}
We can determine the gap between the two lowest Landau bands by $m$ with $k_z=0$.

Next, we will concentrate on an electron-doped quantum limit, namely, the Fermi energy intersects only with the $0+$ Landau band, of which the eigenstate is
\begin{eqnarray}\label{Eq:LB0}
|0,+,k_x,k_z\rangle &=&\begin{bmatrix}
0\\
-i \sin (\theta/2)\\
0\\
\cos(\theta/2)
\end{bmatrix}
|0,k_x,k_z\rangle,
\end{eqnarray}
where we have defined
\begin{eqnarray}
\cos\theta \equiv \frac{-m}{\sqrt{m^2 + (V_n k_z)^2}} ,
\end{eqnarray}
and $|0,k_x,k_z\rangle $ is the state of an usual zeroth Landau level multiplying a plane wave function along the $z$ direction \cite{Lu15prb-QL}.

\begin{figure}
  \centering
\includegraphics[width=0.47\textwidth]{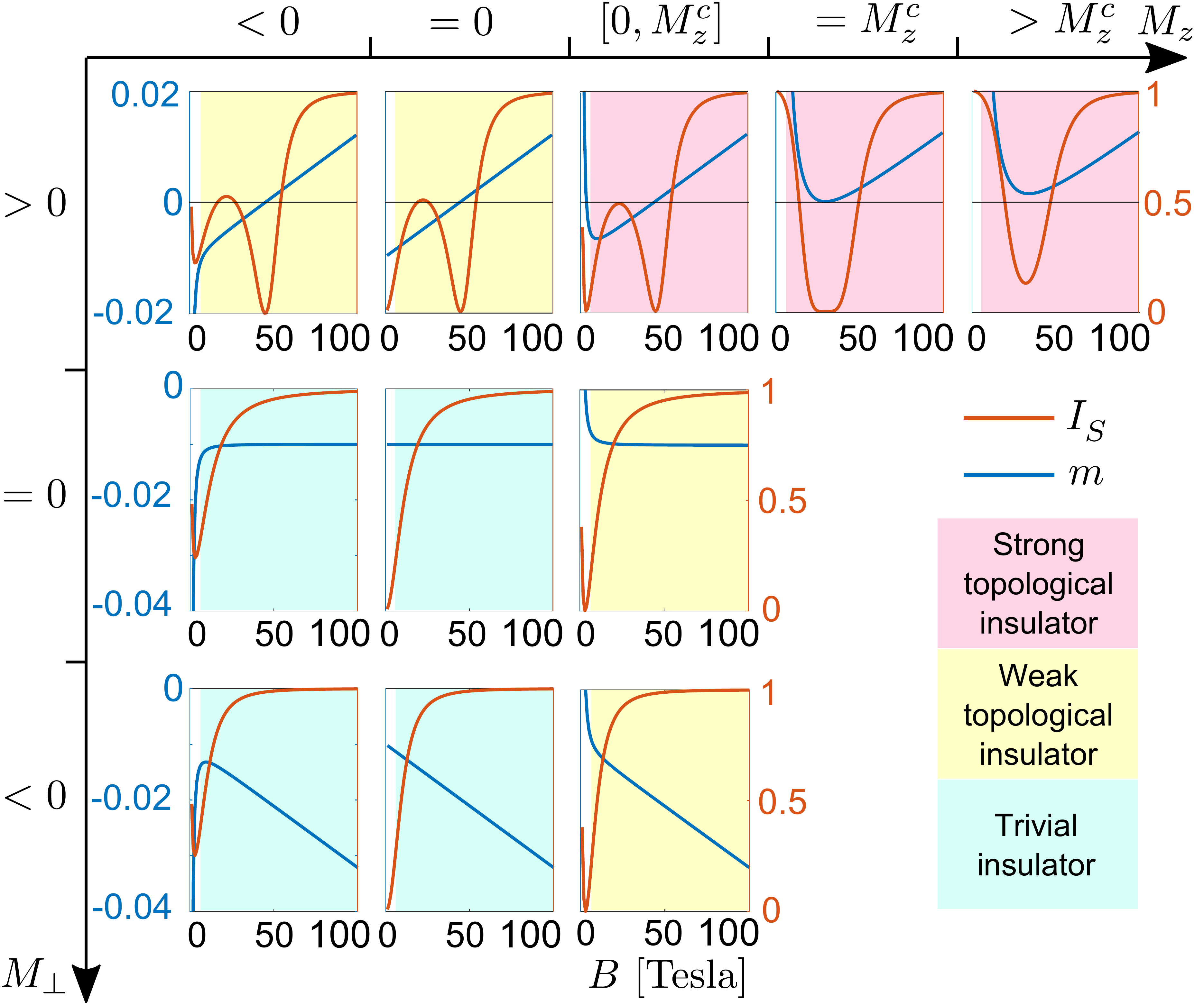}
  \caption{Adapted from Ref. \cite{ChenYY18prl}. The mass term $m$ and form factor $I_S$ as functions of the magnetic field $B$ for $M_0<0$ and different $M_\bot$ and $M_z$. Red, yellow, and green backgrounds indicate the quantum limit for a carrier density of 6$\times$10$^{16}$/cm$^{3}$. Without loss of generality, we have assumed $M_0<0$, so $M_\bot>0$ and $M_z>0$ means strong a topological insulator, $M_\bot\le0$ and $M_z\le 0$ means a trivial insulator, $M_\bot\le0$ and $M_z>0$ means a weak topological insulator (001) and $M_\bot>0$ and $M_z\le0$ means a weak topological insulator (110). The parameters are $M_0$=-0.01 eV, $M_z$=-27, 0, 27, 820, 1200 eV\AA$^2$ from left to right, $M_\bot$=-13.5, 0, 13.5 eV\AA$^2$ from bottom to top. }\label{Fig:Mz}
\end{figure}

 In solids, the electronic transport is relatively influenced by the backscattering, which plays a dictating role in the presence of the 1D Landau band, since there are only two states at the Fermi energy, as denoted by $k_F$ and $-k_F$ in Fig. \ref{Fig:backscattering}. The backscattering between these two states is characterized by the scattering matrix element between them. From the spinor eigenstate in Eq. (\ref{Eq:LB0}), we find that the modular square of the scattering matrix element between the $k_F$ and $-k_F$ states is in proportion to the form factor
\begin{eqnarray}\label{Eq:IS}
 I_S=\left.\cos^2\theta\right|_{k_z=k_F}.
\end{eqnarray}
$I_S$ vanishes when $m=0$, that is, the backscattering between state $k_F$ and state $-k_F$ is forbidden. According to Eq. (\ref{Eq:m}), $m$ vanishes at a critical magnetic field $B_c$  evaluated by
$M_0+M_zk_F^2+ M_\bot eB_c /\hbar=0$,
where $k_z$ is equal to the Fermi wave vector $k_F$ at the Fermi energy.
For a topological insulator, $M_0M_z<0$ and $M_0M_\bot <0$, thus $B_c$ holds finite solutions, at which the backscattering is totally suppressed.
A rich phase diagram can be found in Fig. \ref{Fig:Mz}.
This forbidden backscattering will result in a dip in the resistance as a function of the magnetic field, which can be probed in experiments and can serve as a signature for topological insulator phases. This forbidden backscattering is an eigenstate property and thus is new and different from the mechanism of Landau level crossing \cite{Konig07sci,Buttner11np}, which is a spectrum property.

\subsection{Conductivity in the quantum limit}

Along the direction of the magnetic field, there is no Hall effect, thus the resistivity is the inverse of the conductivity, namely,
$\rho_{zz}=1/\sigma_{zz}$.
In the quantum limit, only band  $0+$ contributes to the conductivity. Following the methods in Refs. \cite{Lu15prb-QL,ZhangSB16njp,Lu11prl},
the resitivitity can be expressed as
\begin{eqnarray}\label{Eq:sigma_zz-form}
  \rho_{zz}=I_S /\sigma_0,
\end{eqnarray}
where $\sigma_0$ is the conductivity independent of the spinor inner product part and $I_S$ is the form factor. For different types of scattering potential, $\sigma_0$ takes different forms. However,  the form factor $I_S$ in Eq. (\ref{Eq:sigma_zz-form}) indicates that, for a topological insulator,  the resistance always has a dip, regardless of $\sigma_0$.
Figure 4 in Ref. \cite{ChenYY18prl} reveals the resistivity when the Gaussian and screened Coulomb potentials are present.
They both show clear dips in the resistivity for some weak topological phases ($M_z\le 0$ and $\widetilde{M}_\bot>0$ in row 1, columns 1 and 2) and strong topological insulator phases ($M_z\in[0,M_z^c]$ and $\widetilde{M}_\bot>0$ in row 1, column 3). Furthermore, the positions of the minima on the $B$ axis do not change for various potentials.
The spin-orbit scattering can improve the above picture and result in a better fitting to the experiment. Specifically, in Fig. \ref{Fig:backscattering} (d), the spin-orbit coupling is included in the screened Coulomb scattering potential. By choosing proper fitting parameters, we obtain a $-40$\% change at the dip of the resistance, which is in agreement with the experiment, as shown in Figs. \ref{Fig:backscattering} (c) and \ref{Fig:backscattering} (d).

\section{Remarks and Perspective}\label{Sec:remark}

The theories of the quantum oscillations need to be improved to match the semiclassical argument and full quantum mechanics calculations. Recently, a quantum theory of intrinsic magnetoresistance for three-dimensional Dirac fermions in a uniform magnetic field is proposed, which shows that the relative magneto-resistance is inversely quartic of the Fermi wave vector and only determined by carrier density, and a formula for the phase shift in SdH oscillation is present as a function of the mobility and the magnetic field \cite{WangHW18prb}. Furthermore, new discoveries on quantization rules in oscillations have been found for graphene, 2D materials, topological metals, topological crystalline insulator, and Dirac and Weyl semimetals \cite{Alexandradinata17prl,Alexandradinata18prx}. Topological contributions have also been found in Bloch oscillations \cite{Holler18prb}. Generalizations of these classical notions to nodal-line systems will be topics of fundamental interests in the future. Whether the rules for phase shift can be generalized to extremal orbits shared by electron and hole pockets in type-II Weyl semimetals \cite{Alexandradinata17prl,OBrien16prl} will be an outstanding problem. Quantum oscillations in type-II Dirac semimetals PdTe$_2$ \cite{Fei2017prb} and nodal-line systems\cite{Yang18prbrc,Oroszlany18prb} have also been addressed.

Lately, the quantized Hall resistance plateaus have been experimentally observed in the topological semimetal Cd$_3$As$_2$ \cite{Uchida17nc,ZhangC17nc-QHE,Schumann18prl}, with thickness ranging from 10 to 80 nm. They cannot be regarded as 2D.
Nevertheless, several questions still hold. First, Cd$_3$As$_2$ is a Dirac semimetal, composed of two time-reversed Weyl semimetals. At a single surface, there is a complete 2D electron gas, formed by two time-reversed half 2D electron gases of the Fermi-arc surface states. There may be also the trivial quantum Hall effect on a single surface. Second, the 3D bulk states quantize into 2D subbands for those thicknesses. If the 3D bulk states cannot be depleted entirely, they also have the trivial quantum Hall effect. The two issues may explain the 2-fold and 4-fold degenerate Hall resistance plateaus observed in the experiments. To deplete the 3D bulk states, the Fermi energy has to be placed exactly at the Weyl nodes. How to distinguish these trivial mechanisms from the 3D quantum Hall effect will be an interesting direction.
Previously, when studying the geometric phase, the parameter space is usually either in real space or momentum space \cite{Xiao10rmp}. The Weyl orbit formed by the Fermi arcs and Weyl nodes is a new physics, because part of the geometric phase is accumulated in real space and part in momentum space, quite different from the parameter spaces studied before. In particular, the geometric phase has a thickness dependence when accumulated along the path as electrons tunnel between the top and bottom surfaces \cite{Potter14nc,ZhangY16srep}. Recently, a new experiment uses this thickness-dependent phase shift to demonstrate the contribution of the Weyl orbit in the observed quantized Hall resistance \cite{ZhangC18nat}. The 3D quantum Hall effect can also be supported by the CDW mechanism \cite{Halperin87jjap}, which has been observed recently in ZrTe$_5$ \cite{Tang18arXiv}.
More works will be inspired to verify the mechanism and realize the 3D quantum Hall effect in the future \cite{Lu18nsr}.

In the quantum limit, our theory have shown that up to two resistance dips may appear if the system is a 3D strong topological insulator \cite{ChenYY18prl}. Surprisingly, recent experiments report up to five oscillations in the quantum limit \cite{WangHC18sa,WangHC18arXiv}. The oscillation as a function of the magnetic field follows a logithimic scale invariance law, much like those in the Efimov bound states of cold atoms.
The Efimov bound state is a three-body bound state arising from the two-body interactions between atoms. Efimov-like bound states have been used to understand the unexpected oscillations \cite{ZhangPF18fop,LiuHW18arXiv}. A direct fitting of the resistance in the experiment have also been demonstrated \cite{WangHC18sa,LiuHW18arXiv,WangHC18arXiv}. Nevertheless, other mechanisms that may lead to the scale invariance oscillations will be topics of broad interest.

In solids, the space group can protect energy nodes with other degeneracies, such as three-, six- and eight-fold one, which may lead to massless fermions that have no counterpart particles in high-energy physics \cite{WengHM16prbrc,WengHM16prb,Bradlyn16sci,Chang2017sr}. Triply-degenerate nodal-point semimetals have been proposed with symmorphic space group symmetry of WC type crystal structure, including TaN, ZrTe and MoP, and observed by angle-resolved photoemission spectroscopy \cite{Lv17nat,MaJZ18np}. The unconventional three component fermions in them are formed by crossing of nondegenerate and double degenerate bands, protected by both rotational and mirror symmetries. As an intermediate fermion between Dirac and Weyl fermion, the host semimetal is expected to have different magnetoresistance \cite{HeJB17prb,ZhuWL17arXiv,Shekhar17arXiv}.
More topics will also be of broad interest, including the quantum transport in magnetic Weyl semimetals \cite{Felser16nmat,Nayake16sa,WangZJ16prl,ChangGQ16srep,NieSM17pnas,YangH17njp,LiuEK18np,WangQ18nc,Guin18arXiv,Chang2018prb_magnetic}, the kagome ferromagnet Fe$_{3}$Sn$_{2}$ \cite{Yin2018nat}, double-Weyl semimetals \cite{Xu11prl,Huang2016pnas}, type-II Weyl semimetals
\cite{Soluyanov15nat,Deng16np,JiangJ17nc,WangYJ16nc,ZhangEZ17nl,ChenD16prb,Khim16prb,Xu2017sciadv,Belopolski2016nc,Belopolski2016prb-TaAs,Change2016sciadv}, hopf-link nodal-line semimetals \cite{ZhongCY17nc,ChenW17prbrc,YanZB17prbrc,Ezawa17prbrc}, Aharonov-Bohm effect \cite{WangLX16nc}, quasiparticle interference on the surfaces of Weyl semimetals \cite{Zheng2016prl,Zheng2018,Zheng2018prl,Chang2016prl-TaAs} and Fano effect \cite{WangS18prl}, exploring axial-gravitational anomaly through thermoelectrical transport \cite{Gooth17nat}.

\acknowledgments
This work was supported by Guangdong Innovative and Entrepreneurial Research Team Program (Grant No. 2016ZT06D348), the National Key R \& D Program (Grant No. 2016YFA0301700), the
National Natural Science Foundation of China (Grant No. 11574127), and the Science, Technology, and Innovation Commission of Shenzhen Municipality (Grant No. ZDSYS20170303165926217 and JCYJ20170412152620376).


%

\end{document}